\documentclass[11pt]{article}
\usepackage[margin=1in]{geometry}
\usepackage[english]{babel}
\usepackage{xr}
\usepackage{setspace}
\usepackage{booktabs}
\usepackage{float}
\usepackage{siunitx}
\usepackage{url}
\usepackage{tikz}
\usetikzlibrary{arrows,automata,positioning}
\usepackage{pgfplots}

\usepackage{mathrsfs,hyperref}
\usepackage[]{amsmath}
\usepackage{amsthm}
\usepackage{fix-cm}
\usepackage[]{amssymb}
\usepackage[]{latexsym}
\usepackage[latin1]{inputenc}
\usepackage[right]{eurosym}
\usepackage[T1]{fontenc}
\usepackage[]{graphicx}
\usepackage[]{epsfig}
\usepackage{fancyhdr}
\usepackage{bbm}
\usepackage{multirow}
\usepackage{natbib}
\usepackage{subcaption}

\allowdisplaybreaks

\makeatletter

\newcommand{\bbr}{\mathbb{R}}

\newcommand{\dcal}{\mathcal{D}}

\newcommand{\scal}{\mathcal{S}}
\newcommand{\ncal}{\mathcal{N}}
\newcommand{\lcal}{\mathcal{L}}

\newcounter{modcount}
\newcommand{\modulo}[2]{%
\setcounter{modcount}{#1}\relax
\ifnum\value{modcount}<#2\relax
\else\relax
\addtocounter{modcount}{-#2}\relax
\modulo{\value{modcount}}{#2}\relax
\fi}
\newcommand{\tablepictures}[4][c]{\begin{tabular}[#1]{@{}c@{}}#2\vspace{0.5cm}\\(\alph{#4}) #3\end{tabular}}
\newcounter{gridsearch}
\newcommand{\tabpic}[2]{
    \stepcounter{gridsearch}
    \modulo{\thegridsearch}{2}
    \ifnum\value{modcount}=0
        \tablepictures[t]{#1}{#2}{gridsearch}\\[2.0cm]
    \else
        \tablepictures[t]{#1}{#2}{gridsearch}&~&
    \fi
}

\makeatother
\newtheorem{lemma}{Lemma}[section]
\newtheorem{proposition}[lemma]{Proposition}
\newtheorem{theorem}[lemma]{Theorem}
\newtheorem{corollary}[lemma]{Corollary}

\newtheorem{example1}[lemma]{Example}
\newtheorem{rem1}[lemma]{Remark}
\newtheorem{assumption}[lemma]{Assumption}
\newtheorem{alg1}[lemma]{Algorithm}
\newtheorem{me1}[lemma]{Mechanism}

\newenvironment{remark}{\begin{rem1}\rm}{\end{rem1}}
\newenvironment{example}{\begin{example1}\rm}{\end{example1}}

\usepackage{color}

\newcommand{\T}{\top}
\newcommand{\diag}{\operatorname{diag}}

\DeclareMathOperator*{\argmax}{arg\,max}

\newcommand\ind[1]{\mathbb{I}_{\{#1\}}}
\newcommand\pdv[2]{\frac{\partial#1}{\partial#2}}
\newcommand{\D}{\mathbb{D}}

\begin{document}

\title{Price mediated contagion through capital ratio requirements with VWAP liquidation prices}
\author{Tathagata Banerjee \thanks{Washington University in St.\ Louis, Department of Electrical and Systems Engineering, St.\ Louis, MO 63130, USA.} \and
Zachary Feinstein \thanks{\textbf{Corresponding author.} Stevens Institute of Technology, School of Business, Hoboken, NJ 07030, USA. \tt{zfeinste@stevens.edu}}}
\date{\today}
\maketitle
\abstract{
We develop a framework for price-mediated contagion in financial systems where banks are forced to liquidate assets to satisfy a risk-weight based capital adequacy requirement. In constructing this modeling framework, we introduce a two-tier pricing structure: the \emph{volume weighted average price} that is obtained by any bank liquidating assets and the \emph{terminal mark-to-market price} used to account for all assets held at the end of the clearing process.  We consider the case of multiple illiquid assets and develop conditions for the existence and uniqueness of clearing prices. We provide a closed-form representation for the sensitivity of these clearing prices to the system parameters, and use this result to quantify: (1) the cost of regulation, in stress scenarios, faced by the system as a whole and the individual banks, and (2) the value of providing bailouts to consider when such notions are financially advisable. Numerical case studies are provided to study the application of this model to data.
}

\section{Introduction}\label{sec:intro}
The modern day financial system is a highly interconnected network. The connections that exist within this network are varied, myriad and complex. These connections might be through direct channels such as interbank debt linkages or through indirect channels such as overlapping portfolios. These connections provide avenues of contagion in the financial networks. Thus negative actions of a bank may cause distress to other firms and eventually affect the entire system. The risk to the financial system, posed by such events, is often called systemic risk. It is imperative that we study and understand how this contagion spreads through financial networks in order to prevent and mitigate systemic crises, e.g., the 2007-2009 financial crisis.

In this work, we study contagion in financial systems from fire sale spillovers. These crises originate when a firm is forced to liquidate its assets to meet some obligation or satisfy some regulation. As firms hold overlapping portfolios this causes impacts globally to all other firms from mark-to-market accounting. These firms are now, themselves, forced to liquidate their assets, which exacerbates the crisis by depressing asset prices further. An important factor in the origin of fire sales is the  unintended consequence of capital regulations in the form of capital or leverage ratios. Due to these regulatory constraints, banks might be forced to delever, setting off a vicious cycle of contagion due to the pro-cyclical nature of these regulatory environments. Such deleveraging occurred in a large scale in the 2008 financial crisis, resulting in amplification of losses. For further discussion on such mechanisms see \cite{BW19,CS19}.

The literature in the study of fire sales may be broadly divided into two different bodies depending on the focus of the study.
The first places more emphasis on the development of a general mathematical framework and exploring questions about, e.g., existence and uniqueness of clearing solutions. Among these works, \cite{CFS05} considers the liquidation problem in the context of a capital adequacy ratio. \cite{AFM16,feinstein2015illiquid} study the fire sale problem when banks are forced to liquidate assets to meet debt obligations. \cite{feinstein2016leverage} develops an extension to \cite{feinstein2015illiquid} where banks, in addition to meeting their debt obligations, must satisfy a leverage ratio. \cite{BW19} considers the problem where banks are required to satisfy a risk-weighted capital ratio. \cite{feinstein2019leverage} extends \cite{BW19} by considering the price-mediated contagion problem in a continuous-time setting to provide results on existence and uniqueness.
The second, broad, body of work on fire sales and price-mediated contagion focuses primarily on the development of an operational modeling framework and the design of stress tests. Typically these results depend on a particular liquidation strategy (e.g., proportional liquidation) and linear price impacts. Some of the notable works in this domain include \cite{GLT15,DE18,CS19}.

In this work, we seek to develop a stress testing framework for price mediated contagion that allows for analytical tractability while also more closely matching the structure of the financial system.
The primary innovation of this work that facilitates the mathematical results is the consideration of both \emph{mark-to-market prices} and \emph{volume weighted average prices} to characterize the system.  In this model, assets are liquidated at the volume weighted average price, but unsold assets are marked to the market.  This distinction between prices was hinted at by~\cite{CS17}, but the implications of that distinction were not fully considered in that work.  By considering this pair of prices we are able to determine conditions for existence and uniqueness under capital adequacy and leverage requirements in equilibrium; as far as the authors are aware, this is the first time that uniqueness results are provided in such a setting.  Similar to the dynamic setting of~\cite{feinstein2019leverage}, the uniqueness results of this work can be used to calibrate the risk-weights to the liquidity of assets; this is in contrast with the prevailing heuristic methodology used in practice.
In addition, we introduce sensitivity results in the fire sale framework which we use to determine two vital statistics: the cost of regulation during a crisis and the value of a rescue fund (whether a bailout external to the system or from other firms in the system).  Sensitivity in the fire sale setting of~\cite{AFM16} was proposed in~\cite{CLY14} to study similar questions; as far as the authors are aware, this work is the first time such results are available in the literature based on leverage-type constraints.
Importantly, all results presented herein hold in markets with multiple assets subject to general price impact functions and with general liquidation strategies satisfying simple, financially meaningful, conditions.
The mathematical results of this work lead to a number of financial implications, namely we are able to determine valid risk-weights for illiquid assets as a mapping of the illiquidity (Corollary~\ref{cor:unique}) and, separately, provide a toolbox to analytically study other policy questions, e.g., the value of a rescue fund (Section~\ref{sec:bailout}).  Furthermore, the general liquidation functions considered herein allow us to compare the proportional liquidation strategy prevalent in the literature (see, e.g.,~\cite{GLT15,CS19,CW19}) to more sophisticated optimization-based strategies; in particular, we find that the tradeoffs typically reported between diversification of the initial balance sheet and diversity of investments no longer hold under this more sophisticated liquidation strategy (Section~\ref{sec:div}).

The remainder of this paper is organized as follows: In Section \ref{sec:setting}, we develop the financial setting for our model and consider liquidation strategies.  Notably, we characterize the liquidation strategies as functions of both the mark-to-market and volume weighted average prices. In Section \ref{sec:clearing}, we formulate the clearing liquidations as a fixed point problem and develop conditions for the existence and uniqueness of these equilibrium prices. In Section \ref{sec:sensitivity}, we formulate the sensitivity analysis of the equilibrium prices with respect to the system parameters as a fixed point problem and provide a closed-form representation for this result. Further, we utilize these results to develop a methodology to evaluate the cost of regulation in stress scenarios and study the value of bailouts. Numerical case studies highlighting the applications of this model are presented in Section \ref{sec:case}.  The proofs of the main results are provided in an Online Appendix.

\section{Financial setting}\label{sec:setting}
We begin with some simple notation that will be consistent for the entirety of this paper.  Let $x,y \in \bbr^n$ for some positive integer $n$, then
\[x \wedge y = \left(\min(x_1,y_1),\min(x_2,y_2),\ldots,\min(x_n,y_n)\right)^\T,\] $x^- = -(x \wedge \vec{0})$ for zero vector $\vec{0} \in \bbr^n$, and $x^+ = (-x)^-$.  Further, to ease notation, we will denote $[x,y] := [x_1,y_1] \times [x_2,y_2] \times \ldots \times [x_n,y_n] \subseteq \bbr^n$ to be the $n$-dimensional compact interval for $y - x \in \bbr^n_+$.  Similarly, we will consider $x \leq y$ if and only if $y - x \in \bbr^n_+$.  In this way we also define monotonicity of vector-valued and multivariate functions in the component-wise sense.

\subsection{Price impacts}\label{sec:price}
For our modeling, we note that banks hold both liquid and illiquid assets.
In line with \cite{BW19,feinstein2019leverage}, we consider two classes of illiquid assets: marketable (stocks or bonds issued by a non-financial corporation) or non-marketable (loans). The distinction between these two classes of illiquid assets is that non-marketable assets are difficult to sell in the short-run, and hence cannot be liquidated during the crisis we consider herein. For further discussion on non-marketable assets see \cite{BW19,DR11}.
Throughout this work we consider a single liquid asset, $m \geq 1$ (marketable) illiquid assets, and an arbitrary number of non-marketable assets.  Each marketable illiquid asset $k$ has $M_k > 0$ outstanding shares; we will denote the vector of outstanding shares by $M \in \bbr^m_{++}$.
Any marketable illiquid asset, when sold, is subject to price impacts.  These price impacts measure the liquidity of an asset; the more liquid an asset the less the prices are affected by market behavior.

With this in mind, we consider the market prices (with the liquid asset acting as the num\'eraire) modeled by an \emph{inverse demand function}.
That is, the function $F: [\vec{0},M] \to \bbr^m_+$ which maps units of illiquid assets being sold into corresponding prices quoted in the market.  Thus we are able to provide the terminal \emph{mark-to-market prices} [MTMP] as a function of the units of assets being sold.
As we focus on financial crises in this work, we consider only short-term pricing and thus do not separately model temporary and permanent price impacts.

Furthermore, we could, instead, consider the demand curve for the nonbanking sector (as done in, e.g.,~\cite{CL15}) which may result from, e.g., an optimization strategy; the inverse demand function could then be constructed from this demand curve.
In such a way, these inverse demand functions $F$ are truly the inverse of the \emph{external} demand for a liquidated portfolio.  In this context, let $D: \bbr^m_+ \to \bbr^m_+$ denote the demand function so that $D(q)$ denotes the demand for the marketable illiquid assets given prices of $q \in \bbr^m_+$.  The inverse demand function is such that $q = F(D(q))$, i.e., the price of the assets is determined so that the demand is matched by the liquidated supply.  Notably, in perfectly competitive (i.e., liquid) markets, these inverse demand functions will provide a constant price for any quantity being liquidated; in this work, we typically assume that the markets are imperfectly competitive and thus liquidations lead to price impacts.  Notably, as can be seen in the formulations given thus far, the price for one asset $F_k(\Gamma)$ can depend on the entire liquidated portfolio $\Gamma \in [\vec{0},M]$ and not just on the amount of asset $k$ sold.  In Assumption~\ref{ass:idf} below, we assume that there are no direct cross impacts, i.e., the price of asset $k$ only depends on the quantity of that asset being sold.

\begin{assumption}\label{ass:idf}
The MTMP exhibits no direct cross impacts, i.e., $F(\Gamma) := \left(f_1(\Gamma_1),\cdots,f_m(\Gamma_m)\right)^\T$ for any $\Gamma \in [\vec{0},M]$.  Additionally, $f_k: [0,M_k] \to (0,1]$ is nonincreasing with $f_k(0) = 1$ for every asset $k = 1,...,m$.  This implies that $M_k$ denotes, also, the original market capitalization for each asset $k$.
\end{assumption}
Assumption~\ref{ass:idf} specifies that our MTMP function $F$ is nonincreasing in liquidations, i.e., prices drop as assets are sold.  This notion is consistent with the construction of the inverse demand function from the demand function $D$ expressed above.  Though we assume that the MTMP exhibits no direct cross impacts, \emph{indirect} cross impacts will be prominent in this work due to the portfolio rebalancing undertaken by market participants. This is highlighted, particularly, in the sensitivity analysis of Section~\ref{sec:sensitivity}.  We wish to note that inverse demand functions are typically assumed to be continuous (see, e.g.,~\cite{AFM16} and Example~\ref{ex:idf}\eqref{ex:idf-lin1}-\eqref{ex:idf-exp}), but we do not restrict ourselves to that condition herein; such a continuity assumption inherently is linked to the notion that marketable assets are infinitely divisible.

However, when liquidating assets, the price attained in the market is a different (higher) price than the MTMP.  Remaining with the notion of a static framework common in the literature, e.g., in~\cite{CFS05,AFM16,feinstein2015illiquid}, we consider this price as the average MTMP throughout the liquidations.  That is, we define the \emph{volume weighted average price} [VWAP] as the mapping $\bar F: [\vec{0},M] \to \bbr^m_+$ such that
\[\Gamma \in [\vec{0},M] \mapsto \bar F_k(\Gamma) := \bar f_k(\Gamma_k) := \frac{1}{\Gamma_k} \int_0^{\Gamma_k} f_k(\gamma_k) d\gamma_k\]
(with $\bar f_k(0) = 1$) for every asset $k$.
Intuitively, the VWAP encodes the notion that liquidating assets captures the entire price history as prices fall (as modeled by the inverse demand function $F$), e.g., the first marginal unit of asset $k$ is sold at the initial MTMP $f_k(0) = 1$ and, after $\gamma_k$ units of asset $k$ have been sold, the next marginal unit is liquidated at the MTMP $f_k(\gamma_k)$.
This distinction between MTMP and VWAP was presented previously in~\cite{AS08} in a single-firm liquidity setting under the terminology ``marginal supply-demand curve'' and ``supply-demand curve'' respectively for the construction of liquidity-based risk measures.
Thus considering these two separate prices makes this model more realistic and enables us to encode a dynamic notion which is absent in the existing systemic risk literature. 
Notably, the VWAP naturally satisfies the property that the value obtained from liquidating assets grows as more assets are sold, i.e., $\Gamma_k \in [0,M_k] \mapsto \Gamma_k \bar f_k(\Gamma_k)$ is strictly increasing.  This property is the primary one introduced in~\cite{AFM16}.

\begin{remark}\label{rem:vwap}
We wish to note that a similar notion of differentiating between the MTMP and VWAP is introduced in~\cite{CS17}.  That work considers a heuristic parameter $\eta \in [0,1]$, which defines the VWAP by $\bar f_k(\Gamma_k) = (1-\eta) + \eta f_k(\Gamma_k)$.  In contrast, we use the MTMP in order to find the actualized VWAP.
\end{remark}

We conclude our discussion of the inverse demand function by considering 4 example functions and the resultant VWAPs. These example inverse demand functions provide explicit parameterizations of the price impacts via, e.g., the market depth.
\begin{example}\label{ex:idf}
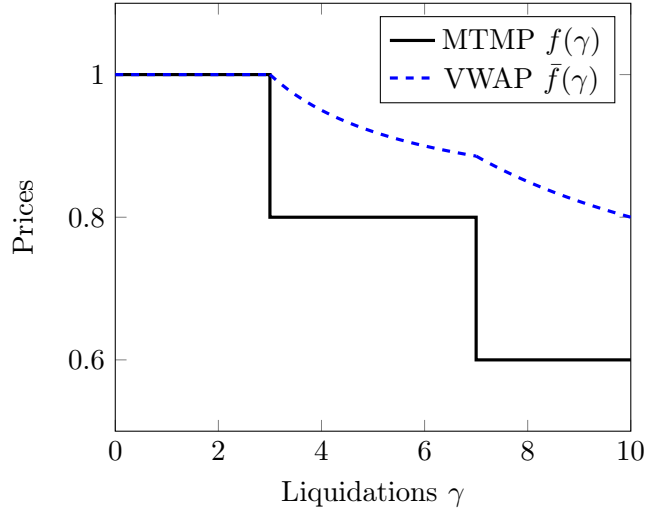
\begin{figure}
\centering
\begin{tikzpicture}
\begin{axis}[
    xlabel={Liquidations $\gamma$},
    ylabel={Prices},
    xmin=0, xmax=10,
    ymin=0.5, ymax=1.1,
    legend pos=north east,
    ]
\addplot [color=black,very thick] coordinates {
    (0,1)
    (3,1)
    (3,0.8)
    (7,0.8)
    (7,0.6)
    (10,0.6)};
\addlegendentry{MTMP $f(\gamma)$};
\addplot [domain=0:3,dashed,color=blue,very thick]{1};
\addplot [domain=3:7,dashed,color=blue,very thick]{(3+.8*(x-3))/x};
\addplot [domain=7:10,dashed,color=blue,very thick]{(3+.8*(7-3)+.6*(x-7))/x};
\addlegendentry{VWAP $\bar f(\gamma)$};
\end{axis}
\end{tikzpicture}
\caption{Illustration of the terminal mark-to-market price and associated volume weighted average price from a limit order book construction as in Example~\ref{ex:idf}\eqref{ex:idf-lob}.}
\label{fig:lob}
\end{figure}

Throughout these examples, and this work, we use the convention that $0/0 = 1$.
\begin{enumerate}
\item\label{ex:idf-lob} First, let us consider a construction of price impacts that follow from a limit order book construction.  In the limit order book, there are fixed price levels with limited liquidity at each level.  Consider price levels $q_{kj} > 0$ with liquidity $m_{kj} > 0$ such that $q_{k,j+1} < q_{kj}$ for every $j \geq 0$ with $q_{k0} = 1$ and $m_{k0} = 0$.  Then the MTMP is defined by $\gamma_k \in [0,M_k] \mapsto \sum_{j \geq 0} q_{kj} \ind{\gamma_k \in [\sum_{i = 0}^j m_{ki} \; , \; \sum_{i = 0}^{j+1} m_{ki})}$.  The resultant VWAP is then given by $\bar f_k(\gamma_k) = \frac{1}{\gamma_k}\sum_{j \geq 0} q_{kj} \left[\gamma_k \wedge \sum_{i = 0}^{j+1} m_{ki} - \gamma_k \wedge \sum_{i = 0}^j m_{ki}\right]$ for $\gamma_k \in [0,M_k]$.  An example of this inverse demand function setting is provided in Figure~\ref{fig:lob}.
\item\label{ex:idf-lin1} Second, consider a generalization of the linear price impact function common in the literature (e.g., \cite{BW19,DE18,CS19,GLT15}).  Let the MTMP be defined by $\gamma_k \in [0,M_k] \mapsto f_k(\gamma_k) = 1 - b \gamma_k^a$ for $a \geq 0$ and $b \in [0,M_k^{-\alpha})$.  The resultant VWAP is then given by $\bar f_k(\gamma_k) = 1 - \frac{b}{1 + a} \gamma_k^a$ for any $\gamma_k \in [0,M_k]$.  In particular, this satisfies the construction of the VWAP from \cite{CS17} with $\eta = \frac{1}{1+a}$ as described in Remark~\ref{rem:vwap}.  We wish to note that many commonly used inverse demand functions exist in this setting: if $a = 1$ then this is the linear inverse demand function studied in, e.g., \cite{BW19,DE18,CS17,GLT15} and utilized in the case studies of Section~\ref{sec:case}; if $a = \frac{1}{2}$ then this satisfies the square root law for market impacts as considered in, e.g., \cite{PRST2018dimensional} and citations therein; if $a = 2$ then this has a quadratic price impact as in, e.g., \cite{almgrenchriss2001}.
\item\label{ex:idf-lin2} Third, consider a different generalization of the linear price impact function.  Let the MTMP be defined by $\gamma_k \in [0,M_k] \mapsto f_k(\gamma_k) = (1 - b \gamma_k)^a$ for $b \leq M_k^{-1}$ with $a b \geq 0$.  The resultant VWAP is then given by $\bar f_k(\gamma_k) = \begin{cases}\frac{1}{(1+a)b\gamma_k}\left[1 - (1 - b\gamma_k)^{1+a}\right] &\text{if } a \neq -1 \\ -\frac{\log(1 - b\gamma_k)}{b\gamma_k} &\text{if } a = -1\end{cases}$ for any $\gamma_k \in [0,M_k]$.
\item\label{ex:idf-exp} Finally, consider the exponential inverse demand function for the MTMP, i,e, $\gamma_k \in [0,M_k] \mapsto f_k(\gamma_k) = \exp(-b \gamma_k)$ for $b \geq 0$.  The resultant VWAP is then given by $\bar f_k(\gamma_k) = \frac{1}{b \gamma_k}\left[1 - \exp(-b \gamma_k)\right]$ for any $\gamma_k \in [0,M_k]$.
\end{enumerate}
\end{example}
Herein we only consider inverse demand functions as a theoretical construction rather than focus on the empirical calibration of such functions.  We refer to, e.g.,~\cite{GLT15} for considerations of the market depth for (linear) inverse demand functions.

\subsection{The stylized balance sheet}\label{sec:bs}
We will consider two time points $t\in\{0,1\}$. At $t=0$, a bank or firm holds $x \geq 0$ liquid assets (e.g., cash). As mentioned in the prior section, we will assume, without loss of generality, that the price of this asset stays constant at $1$ at all times. In addition to this liquid asset, the bank portfolio is comprised of illiquid assets.
We assume that each bank holds $s \in [\vec{0},M] \subseteq \bbr^m_+$ shares of the marketable illiquid assets and $\ell \geq 0$ of non-marketable assets. Without loss of generality, and as in Assumption~\ref{ass:idf}, we assume that the price of all the (marketable) illiquid assets are $F(\vec{0}) = \bar F(\vec{0}) = \vec{1}$ at time $0$.

On the other side of the balance-sheet each bank or firm has $\bar{p} \geq 0$ in liabilities.  For simplicity in this work, we will assume that all liabilities are not held by any other firms in this system; additionally, we will assume that no liabilities come due during the (short time horizon) of the fire sale cascade under study, but are liquid enough that they can frictionlessly be paid off early with liquid assets.  Thus, at time $0$, this bank has a capital of $x+\ell+\vec{1}^\T s-\bar p$. This is depicted in Figure~\ref{fig:balance_sheet}.
\begin{remark}\label{rem:default-contagion}
Prior fire sale literature, e.g.,~\cite{CFS05,AFM16,AW_15} consider the joint impacts of \emph{default contagion} and price-mediated contagion; those works study default contagion by introducing interbank assets and liabilities.  We study the simpler setting with external assets and liabilities only because, empirically, the price mediated contagion has been shown to cause much larger harm to the financial system than default contagion (see, e.g.,~\cite{feinstein2020interbank}).
\end{remark}

\begin{figure}[t]
\centering
\begin{tikzpicture}
\draw[draw=none] (0,6.5) rectangle (6,7) node[pos=.5]{\bf Balance Sheet $t=0$};
\draw[draw=none] (0,6) rectangle (3,6.5) node[pos=.5]{\bf Assets};
\draw[draw=none] (3,6) rectangle (6,6.5) node[pos=.5]{\bf Liabilities};

\filldraw[fill=blue!20!white,draw=black] (0,5) rectangle (3,6) node[pos=.5,style={align=center}]{Liquid \\ $x$};
\filldraw[fill=yellow!20!white,draw=black] (0,1.8) rectangle (3,5) node[pos=.5,style={align=center}]{Illiquid \\ (Marketable) \\ $\vec{1}^\T s$};
\filldraw[fill=green!20!white,draw=black] (0,0) rectangle (3,1.8) node[pos=.5,style={align=center}]{Illiquid \\ (Non-marketable) \\ $\ell$};

\filldraw[fill=purple!20!white,draw=black] (3,3) rectangle (6,6) node[pos=.5,style={align=center}]{Total \\ $\bar p$};
\filldraw[fill=orange!20!white,draw=black] (3,0) rectangle (6,3) node[pos=.5,style={align=center}]{Capital \\ $x+\ell+\vec{1}^\T s$ \\ $-\bar p$};

\draw[->,line width=1mm] (6.5,3) -- (8.5,3);

\draw[draw=none] (9,6.5) rectangle (15,7) node[pos=.5]{\bf Balance Sheet $t=1$};
\draw[draw=none] (9,6) rectangle (12,6.5) node[pos=.5]{\bf Assets};
\draw[draw=none] (12,6) rectangle (15,6.5) node[pos=.5]{\bf Liabilities};

\filldraw[fill=blue!20!white,draw=none] (9,5) rectangle (12,6) node[pos=.5,style={align=center}]{Liquid \\ $x$};
\filldraw[fill=blue!20!white,draw=none,opacity=0.5] (9,4.5) rectangle (12,5) node[pos=.5,style={align=center}]{$\bar{q}^\T \gamma$};
\filldraw[fill=yellow!20!white,draw=none,opacity=0.5] (9,4.5) rectangle (12,5) node[pos=.5,style={align=center}]{$\bar{q}^\T \gamma$};
\draw[draw=none] (9,4.5) rectangle (12,5) node[pos=.5]{$\bar{q}^\T \gamma$};
\draw[dotted] (9,5) -- (12,5);
\filldraw[fill=yellow!20!white,draw=black] (9,2.8) rectangle (12,4.5) node[pos=.5,style={align=center}]{Illiquid \\ (Marketable) \\ $q^\T(s-\gamma)$};
\filldraw[fill=green!20!white,draw=black] (9,1) rectangle (12,2.8) node[pos=.5,style={align=center}]{Illiquid \\ (Non-marketable) \\ $\ell$};
\filldraw[fill=yellow!20!white,draw=black] (9,0) rectangle (12,1);

\filldraw[fill=purple!20!white,draw=black] (12,3) rectangle (15,6) node[pos=.5,style={align=center}]{Total \\ $\bar p$};
\filldraw[fill=orange!20!white,draw=black] (12,1) rectangle (15,3) node[pos=.5,style={align=center}]{Capital \\ $ C := x+\ell+$\\$\bar q^\T \gamma + q^\T [s-\gamma]$\\$-\bar{p}$};
\filldraw[fill=orange!20!white,draw=black] (12,0) rectangle (15,1);
\draw (9,0) rectangle (15,6);
\draw (12,0) -- (12,6);

\begin{scope}
    \clip (9,0) rectangle (15,1);
    \foreach \x in {-9,-8.5,...,15}
    {
        \draw[line width=.5mm] (9+\x,0) -- (15+\x,6);
    }
\end{scope}
\end{tikzpicture}
\caption{Stylized balance sheet for a firm before and after liquidation updates.}
\label{fig:balance_sheet}
\end{figure}

At time $t=1$, a firm or bank may liquidate some of its marketable illiquid assets in order to satisfy regulatory requirements after, e.g., some shock to the balance sheet of the bank such as writedowns on the non-marketable assets $\ell$; this is detailed in much greater detail in Section~\ref{sec:car} below.  In particular, the bank sells $\gamma \in [\vec{0},s]$ at the VWAP $\bar q \in (0,1]^m$ to obtain $\bar q^\T \gamma \geq 0$ liquid assets.  The remaining $s - \gamma$ marketable illiquid assets in the balance sheet are valued at the MTMP $q \in (\vec{0},\bar q]$.  We will denote this space of joint MTMP and VWAP prices by the lattice
\[\D := \{(q,\bar q) \in [F(M),F(\vec{0})] \times [\bar F(M),\bar F(\vec{0})] \; | \; q \leq \bar q\}.\]
The capital at time $t = 1$ is thus given by $C := x + \ell + \bar q^\T \gamma + q^\T [s - \gamma] - \bar p$.  This is depicted in Figure~\ref{fig:balance_sheet}.

\begin{remark}\label{rem:VWAP}
While computing the capital of a bank, the proceeds from liquidation is given by $\bar q^\T \gamma$. This is in contrast to much of the existing literature on fire sales (e.g.,~\cite{BW19,feinstein2016leverage}) where no distinction is made between liquidation price and current price; in prior literature the proceeds would be computed as $q^\T \gamma$. By introducing the MTMP and VWAP prices, this is a more realistic scenario and it offers a better interpretation of the capital of a firm.  If a single price had been considered, the capital of a firm ($x + \ell + q^\T s - \bar p$) would have been independent of liquidations $\gamma$ and would depend only on the price $q$.
\end{remark}

For the remainder of this work we consider $n \geq 1$ banks.  We will denote the set of all banks in the system by $\ncal := \{1,2,...,n\}$.
In terms of vector notation, at time $t=0$, the banks are holding an amount $x \in \bbr^n_+$ of liquid assets, $\ell \in \bbr^n_+$ shares of non-marketable illiquid assets, $S=(s_{ik}) \in \bbr^{n\times m}_+$ shares of  marketable illiquid assets, and an amount $\bar{p} \in \bbr^n_+$ in liabilities.  By construction, we will always set $\sum_{i \in \ncal} s_{ik} \leq M_k$ for all assets $k$.  At time $t=1$, the banks are liquidating $\Gamma=(\gamma_{ik}) \in [{\bf 0},S]$ of the marketable illiquid assets where ${\bf 0}$ denotes the $n \times m$ zero matrix.

%
%
%
%

\subsection{Capital adequacy ratio and liquidation strategies}\label{sec:car}
The Basel regulation mandates the use of a capital adequacy ratio (also called the risk-weighted capital ratio) to assess the solvency of banks. The risk-weighted capital ratio is defined as
\begin{equation*}
    \text{Risk-weighted capital ratio}=\frac{\text{Capital}}{\text{Risk-Weighted Assets}}
\end{equation*}

The determination of the risk-weights of different assets requires the consideration of a number of complex factors.  In this case, we make the assumption that these risk-weights are known to us and given by $0$ for the liquid asset and $\alpha_k \geq 0$ for marketable illiquid asset $k=1,...,m$. For the non-marketable asset we let the risk-weight be dependent on each bank and let $\alpha_{\ell,i} \geq 0$ be the risk-weight for the non-marketable assets of bank $i \in \ncal$. To simplify notation, we define $A := \text{diag}(\alpha_1,...,\alpha_m)$ to be the diagonal matrix of the risk-weights of the marketable illiquid assets.

Thus at time $t=0$, the risk-weighted capital ratio $\theta_i$ of bank $i$ is given by
\begin{equation*}
\theta_i(0) = \frac{x_i +\ell_i +\vec{1}^\T s_i-p_i}{\vec{1}^\T A s_i+\alpha_{\ell,i} \ell_i}
\end{equation*}
According to banking regulations, banks are required to maintain a minimum capital ratio $\theta_{\min} > 0$, e.g., $8\%$ in Basel II regulations.
We wish to note that this capital adequacy ratio is related to a leverage requirement if all risk-weights are set to $1$.

At time $t=0$, banks may or may not be in compliance with this regulatory constraint (the initial balance sheet could be constructed via, e.g., portfolio optimization constrained by the capital regulation; such a construction is outside the scope of this work).  In~\cite{BW19}, it is assumed that an external shock occurs at time $t=0^+$ to which banks must react at time $t=1$; herein, we do not explicitly differentiate between times $t=0$ and $t=0^+$ to simplify notation.  If we were to consider the shock setting, this could be a shock to the value of liquid or illiquid assets (marketable or non-marketable) as is standard in the literature (see, e.g.,~\cite{BW19,GLT15,DE18}) or a shock in the risk-weights, i.e., where the risk-weight of an asset jumps due to, e.g., a credit downgrade. We specifically wish to highlight the setting in which the non-marketable illiquid assets are shocked in order to draw comparisons to the 2007-2009 financial crisis in which the stressors were from subprime mortgages going unpaid (direct shocks leading to a drop in $\ell_i$) and credit downgrades of those same assets (indirect shocks leading to increases in $\alpha_{\ell,i}$).

Depending on the state of the balance sheets of the financial institutions at $t=0$, the capital ratio of some banks may fall below the regulatory minimum $\theta_{\min}$. In this situation, banks typically have two options: issue new equity or liquidate portions of the balance sheet.
At the time of a crisis, issuing new equity might not be feasible. As such, banks will be forced to liquidate assets to meet the regulatory constraint.  However, these liquidations can set off a fire sale causing additional losses to the system.

When a bank is forced to liquidate assets, they do so in a minimal way.  That is, they liquidate the minimal value necessary to recover
\[\theta_i(1) := \frac{x_i + \ell_i + \bar q^\T \gamma_i + q^\T [s_i - \gamma_i] - \bar p_i}{q^\T A [s_i - \gamma_i] + \alpha_{\ell,i}\ell_i} \geq \theta_{\min}\]
and will do nothing if they already satisfy this regulatory requirement.
This practice is in line with the existing literature \cite{BW19,GLT15,DE18,feinstein2019leverage} as well as empirical evidence \cite{AS10}.
It might be entirely possible that even when a bank liquidates all its assets it cannot restore its capital ratio to $\theta_{\min}$ at time $1$. In this situation we will assume that such a bank is insolvent and costlessly liquidated at $t=1$ (while liquidating all of its marketable assets in the market and thus depressing those prices) along the lines of~\cite{BW19}; this costless liquidation of non-marketable assets can be viewed as a consequence of collateralized borrowing as these assets will be transferred, rather than sold, in case of insolvency.

To this end, bank $i \in \ncal$ can belong to any of the following three mutually exclusive and exhaustive sets given the prices $(q,\bar q) \in \D$. (Implicitly this setting assumes banks act as price takers as is commonly studied in, e.g.,~\cite{CFS05,GLT15,AFM16}; the price \emph{making} situation is studied explicitly in Proposition~\ref{prop:equilibrium} below.)  To shorten notation throughout this work we will define the \emph{shortfall} of bank $i$ as $h_i := \bar p_i - x_i - (1 - \alpha_{\ell,i}\theta_{\min})\ell_i$, i.e., the excess liabilities above the liquid assets and non-marketable assets (reduced by the risk-weights).  This shortfall provides the liquidity that needs to be raised via asset liquidations as can be seen immediately below. 
\begin{itemize}
\item \textit{Solvent and liquid}: Let us denote this set of banks that need not liquidate any assets in order to satisfy the capital ratio requirement $\theta_i(1) \geq \theta_{\min}$ by $\scal(q, \bar q)$. For any bank $i \in \scal(q,\bar q)$, $\gamma_i(q,\bar q) = \vec{0}$.  In fact, this set is characterized by $h_i \leq q^\T [I - A\theta_{\min}]s_i$.
\item \textit{Solvent but illiquid}: Let us denote this set of banks that need to liquidate marketable illiquid assets in order to satisfy the capital ratio requirement $\theta_i(1) \geq \theta_{\min}$ by $\lcal(q, \bar q)$. For any bank $i \in \lcal(q,\bar q)$,
\[\left(\bar q - [I - A\theta_{\min}]q\right)^\T \gamma_i(q,\bar q) = h_i - q^\T [I - A\theta_{\min}] s_i.\]  In fact, this set is characterized by $q ^\T [I-A\theta_{\min}]s_i < h_i < \bar q^\T s_i$.
\item \textit{Insolvent}: Let us denote this set of banks that are unable to satisfy the capital ratio requirement $\theta_i(1) < \theta_{\min}$ regardless of asset liquidations by $\dcal(q, \bar q)$. For any bank $i \in \dcal(q,\bar q)$, $\gamma_i(q,\bar q)=s_i$.  In fact, this set is characterized by $h_i \geq \bar q^\T s_i$.
\end{itemize}
These conditions are encoded mathematically in Assumption~\ref{ass:liquidation} below.
We wish to note that throughout this work we define the liquidity of firms via their market liquidity rather than funding liquidity since, as previously mentioned, we assume that issuing of new equity might not be a feasible method to raise funding at the time of a crisis.  Since the 2008 financial crisis, banks are subject to liquidity ratio requirements as well as solvency ratio requirements for exactly such concerns; we utilize the capital adequacy ratio for both solvency and liquidity concerns.
\begin{assumption}\label{ass:liquidation}
Given a coupled MTMP and VWAP $(q,\bar q) \in \D$, the system of banks will liquidate $\Gamma(q,\bar q) \in [{\bf 0},S]$ marketable illiquid assets satisfying the \emph{minimal liquidation condition} for each bank $i$:
\begin{equation}
\label{eq:mlc} \left(\bar q - [I - A\theta_{\min}]q\right)^\T\gamma_i(q,\bar q) = \left(h_i - q^\T [I - A\theta_{\min}] s_i\right)^+ \wedge \left[\left(\bar q - [I - A\theta_{\min}]q\right)^\T s_i\right]
\end{equation}
for shortfall $h_i := \bar p_i - x_i - (1 - \alpha_{\ell,i}\theta_{\min})\ell_i$.
\end{assumption}

We refer to~\eqref{eq:mlc} as the minimal liquidation condition as it provides the minimal amount needed to be liquidated in order to satisfy the capital adequacy requirement $\theta_i(1) \geq \theta_{\min}$, or liquidate all assets if insolvent.
This is similar to the liquidity constraint  used in \cite{feinstein2015illiquid,feinstein2016leverage}.

\begin{remark}
In general, there are some immediate conclusions that can be made on the solvency and liquidity of the different institutions at the coupled MTMP and VWAP $(q,\bar q) \in \D$.  Specifically,
\begin{align*}
\scal(q,\bar q) &\subseteq \scal(\vec{1},\vec{1}) = \{i \in \ncal \; | \; \theta_i(0) \geq \theta_{\min}\}\\
\lcal(q,\bar q) &\subseteq \lcal(\vec{1},\vec{1}) \cup \scal(\vec{1},\vec{1})\\
\dcal(q,\bar q) &\subseteq \dcal(\vec{1},\vec{1}) \cup \lcal(\vec{1},\vec{1}) \cup \scal(\vec{1},\vec{1}) = \ncal.
\end{align*}
That is, any firm is solvent and liquid at prices $(q,\bar q)$ only if it is solvent and liquid at time $t = 0$ (with equality if $q = \bar q = \vec{1}$).  Otherwise, an insolvent firm at time $t = 0$ (i.e., $\theta_i(0) < \theta_{\min}$) may become either solvent but illiquid or insolvent -- but \emph{not} solvent and liquid -- at prices $(q,\bar q)$ due to liquidations and the minimal liquidation condition.  Additionally, as prices drop, the solvency and liquidity of firms similarly decreases.
This is studied in more detail in a small numerical setting in Section~\ref{sec:discussion}.
\end{remark}

Before considering some sample liquidation functions, we wish to give one more consideration of the risk-weights.
\begin{assumption}\label{ass:alphatheta}
Throughout this work we will assume that $\alpha_k \theta_{\min} < 1$ for every asset $k=1,2,...,m$.
\end{assumption}
If $\alpha_k \theta_{\min} \geq 1$ for any $k$, then the setting of this paper implies that as price drops in that asset, the bank will always satisfy the capital regulation which is opposite to the scenario that we are modeling. For further discussion see Remark 2.2 of \cite{feinstein2019leverage}.

We complete this section by discussing a few sample liquidation functions satisfying Assumption~\ref{ass:liquidation}.
\begin{example}\label{ex:gamma}~
\begin{enumerate}
\item\label{ex:gamma-1asset}
Consider a setting with only $m = 1$ marketable illiquid asset.  In this setting,~\eqref{eq:mlc} uniquely defines the liquidations any bank would take:
\[\gamma(q,\bar q) = \left(\frac{h - q(1-\alpha\theta_{\min})s}{\bar q - (1-\alpha\theta_{\min})q}\right)^+ \wedge s\]
for any $(q,\bar q) \in \D$.  We wish to note that this liquidation function $\gamma$ is continuous and nonincreasing in $(q,\bar q) \in \D$.
\item\label{ex:gamma-proportional}
Consider a setting in which a bank wishes to maintain its current portfolio ratio, i.e., when it liquidates it sells shares of its full portfolio. Proportional liquidation has been widely explored in the existing literature (e.g., \cite{DE18,GLT15,CS17}) for the analysis of fire sales.  Such a strategy is defined by:
\[\gamma(q,\bar q) = \left[\left(\frac{h - q^\T[I - A\theta_{\min}]s}{(\bar q - [I - A\theta_{\min}]q)^\T s}\right)^+ \wedge 1\right] s\]
for any $(q,\bar q) \in \D$. We wish to note that this liquidation function $\gamma$ is continuous and nonincreasing in $(q,\bar q) \in \D$.
\item\label{ex:gamma-utility}
Consider now the first of two strategies based on utility maximization.
In this first setting, a bank decides on its liquidation strategy $\gamma$ so as to maximize some strictly concave utility function $u$. As such we consider firms to be utility maximizers of the following problem rather than following a mechanical property:
\[\gamma(q,\bar q) = \argmax_{\gamma \in G(q,\bar q)} u(\gamma)\]
with constraint set provided by a relaxation of the minimal liquidation condition:
\[G(q,\bar q) = \left\{\gamma \in [\vec{0},s] \; | \; (\bar q - [I - A\theta_{\min}]q)^\T \gamma \geq (h - q^\T [I - A\theta_{\min}] s)^+ \wedge [(\bar q - [I - A\theta_{\min}]q)^\T s]\right\}\]
for any $(q,\bar q) \in \D$.
By strict concavity and using Berge's maximum theorem, we can show that this utility maximizing liquidation function is continuous.  If $u$ is also decreasing then this will satisfy the minimal liquidation condition~\eqref{eq:mlc}.
\item\label{ex:gamma-equilibrium} Consider now an extension of the utility maximizer, but one that also depends on the actions of other firms.  Consider bank $i$ to follow this strategy and the aggregate liquidations of all other banks is given by $\gamma_{-i}^* \in \bbr^m_+$:
\[\gamma_i(q,\bar q;\gamma_{-i}^*) = \argmax_{\gamma_i \in G(q,\bar q)} u_i(\gamma_i,\gamma_{-i}^*)\]
for prices $(q,\bar q) \in \D$.  We call this a price taking equilibrium strategy since, if multiple firms follow this strategy, the actualized liquidations would be the solution of a game theoretic problem with given prices $(q,\bar q) \in \D$. Notably, so long as the utility functions are strictly concave then this liquidation strategy is continuous w.r.t.\ the prices and liquidations of the other firms (by Berge's maximum theorem) allowing for the actualized liquidations to be computed as the solution of a fixed point problem.  As with the utility maximizer liquidation function, if $u_i$ is also decreasing then this strategy will satisfy the minimal liquidation condition~\eqref{eq:mlc}.  Furthermore, if all firms utilizing this liquidation strategy construct a diagonally strictly concave game (see, e.g.,~\cite{rosen1965}) then there exists a unique price taking equilibrium liquidation strategy at given prices, and that strategy is continuous in the prices $(q,\bar q) \in \D$.
Finally, as indicated by the nomenclature used, the banks following this strategy -- as with all prior example liquidation functions -- are price takers in the market.  The setting in which banks act as price makers in an equilibrium setting is presented in Proposition~\ref{prop:equilibrium} below.
\end{enumerate}
\end{example}

\section{Clearing formulation and solutions}\label{sec:clearing}
With this thorough discussion of the financial setting, we now wish to consider the problem of finding the clearing prices.  Though the minimal liquidation condition~\eqref{eq:mlc} gives information about the liquidations that bank $i$ performs under the MTMP and VWAP $(q,\bar q) \in \D$, when actualizing these sales the prices will adjust according to the inverse demand functions $F$ and $\bar F$.  Therefore, this problem can accurately be modeled using a fixed point equation.

As discussed above, consider the matrix of liquidations $\Gamma(q,\bar q) \in [{\bf 0},S]$ at MTMP and VWAP of $(q,\bar q) \in \D$.  We seek an equilibrium price so that the resultant price from liquidations is equal to the prices that generate those liquidations.  That is, we seek the fixed point to the function $\Phi: \D \to \D$ defined by:
\begin{equation}\label{eq:clearing}
(q,\bar q) \in \D \mapsto \Phi(q,\bar q) := \left(F(\Gamma(q,\bar q)^\T \vec{1}) \; , \; \bar F(\Gamma(q,\bar q)^\T \vec{1})\right).
\end{equation}
We call $(q,\bar q) \in \D$ a clearing solution or clearing prices if $(q,\bar q) = \Phi(q,\bar q)$.
This is distinct from the algorithmic and iterative approach in which banks act and capture different prices over (potentially fictitious) time, we refer to, e.g.,~\cite{feinstein2019leverage} for such a dynamic setting; we focus on this equilibrium setup for analytical tractability for, e.g., sensitivity analysis considered in Section~\ref{sec:sensitivity}.

In this section, we develop conditions for existence and uniqueness of the clearing prices. In the existing literature on price-mediated contagion due to leverage and capital adequacy ratio requirements, existence of (static) clearing solutions has been explored for the one asset case in~\cite{BW19} and for the multi-asset case in~\cite{feinstein2016leverage}. However in these works, uniqueness of the clearing prices has not been previously solved.  This is tackled directly in Theorem~\ref{thm:unique}.  First, in Proposition~\ref{prop:exist}, we present simple conditions for the existence of the clearing prices.

\begin{proposition}\label{prop:exist}
Consider the regulatory and balance sheet assumptions from Section~\ref{sec:setting}.
\begin{enumerate}
\item\label{prop:exist-brouwer} Let the inverse demand function $F: [\vec{0},M] \to (0,1]^m$ be continuous. If the liquidation function $\Gamma: \D \to [{\bf 0},S]$ is continuous, then there exists a clearing price $(q^*, \bar q^*)$.
\item\label{prop:exist-tarski} If the liquidation function $\Gamma: \D \to [{\bf 0},S]$ is nonincreasing, then there exists a greatest and least clearing price $(q^\uparrow, \bar q^\uparrow) \geq (q^\downarrow, \bar q^\downarrow)$.
\end{enumerate}
\end{proposition}

Now we present the primary theoretical result of this section; that is, we consider the uniqueness of these clearing prices.
The adoption of the VWAP in this framework, besides providing a more realistic financial framework, offers significant mathematical advantages, particularly in the analysis of uniqueness as is evident from the preceding theorem as this was unable to be proven in, e.g.,~\cite{BW19,feinstein2016leverage}.

\begin{theorem}\label{thm:unique}
Consider the setting of Proposition~\ref{prop:exist}\eqref{prop:exist-tarski}.
Let the inverse demand function be such that
\[\Gamma^* \in [\vec{0},M] \mapsto \bar F(\Gamma^*)^\T \Gamma^*+F(\Gamma^*)^\T [I-A\theta_{\min}] (M-\Gamma^*)\]
is strictly increasing.  If the liquidation function $\Gamma: \D \to [{\bf 0},S]$ satisfies Assumption~\ref{ass:liquidation}, then there exists a unique clearing price $(q^*, \bar q^*)$.
\end{theorem}

\begin{remark}
We want to point out the similarity in the uniqueness condition presented in this work to one of the very few uniqueness result in the fire sale literature as presented in~\cite{AFM16,feinstein2015illiquid}. In those works, the liquidation occurs based on a leverage ratio ($\alpha = \vec{1}$) with no leverage allowed ($\theta_{\min} = 1$). Our new condition on the inverse demand function in Theorem~\ref{thm:unique} reduces exactly to the condition from those papers.  In fact, given our construction of the VWAP from the MTMP, the property $\Gamma^* \in [\vec{0},M] \mapsto \bar F(\Gamma^*)^\T \Gamma^*$ strictly increasing is automatically satisfied.  Thus our result can be seen as a generalization of the uniqueness condition provided in~\cite{AFM16,feinstein2015illiquid}.
\end{remark}

Theorem \ref{thm:unique} provides a condition for the uniqueness of solution for an equilibrium price $(q^*, \bar q^*)$ in terms of the inverse demand function $F$. However this condition also depends on the risk-weights $\alpha$. We can make this dependence explicit by stating the uniqueness condition in terms of the inverse demand function $F$ and the risk-weights $\alpha$ along the lines of \cite{feinstein2019leverage}. This is described in the following corollary.  This is important as it allows us to calibrate the risk-weights to the liquidity of the assets.

\begin{corollary}\label{cor:unique}
Let the inverse demand function $F$ be differentiable and such that $\Gamma_k^* \in [0,M_k] \mapsto \frac{(M_k - \Gamma_k^*) f_k'(\Gamma_k^*)}  {f_k(\Gamma_k^*)}$ is nondecreasing for every asset $k=1,2,..,m$. Additionally, let $\alpha_k \in \frac{1}{\theta_{\min}} \times (-\frac{M_k f_k'(0)}{1-M_k f_k'(0)} \; , \; 1)$ for every asset $k$.  If the liquidation function $\Gamma: \D \to [{\bf 0},S]$ is nonincreasing and satisfies Assumption~\ref{ass:liquidation}, then there exists a unique clearing price $(q^*,\bar q^*)$.
\end{corollary}

\begin{remark}\label{rem:unique}
Corollary~\ref{cor:unique} provides sufficient conditions for the uniqueness property given in Theorem~\ref{thm:unique}.  As such, it is a stronger set of conditions than presented in Theorem~\ref{thm:unique}. However, we feel that the conditions of Corollary~\ref{cor:unique} are easier to test and evaluate as it separates the conditions on the inverse demand function and the risk-weights, while also providing a nice interpretation in terms of calibrating risk-weights.  We wish to note that the conditions of Corollary~\ref{cor:unique} exactly coincide with those provided in Lemma 3.11 of \cite{feinstein2019leverage} in a continuous-time model of the proportional liquidation setting.  The condition on the inverse demand function ($\Gamma_k^* \in [0,M_k] \mapsto \frac{(M_k - \Gamma_k^*) f_k'(\Gamma_k^*)}  {f_k(\Gamma_k^*)}$ is nondecreasing) is discussed in detail in Remark~3.5 of that work.  In short, this condition implies that a financial institution does not need to increase the speed it is selling the illiquid assets solely to counteract its own impacts.
\end{remark}

\begin{example}\label{ex:unique-idf}
We now wish to consider our example inverse demand functions to determine under which scenarios they satisfy the differentiability and monotonicity condition of Corollary~\ref{cor:unique} with nonincreasing liquidation functions (e.g., proportional liquidation).  Though the utility maximizing and equilibrium liquidation strategies have clearing solutions, the results of this work are not strong enough to guarantee uniqueness of those clearing prices.
\begin{enumerate}
\item The limit order book setting is not differentiable, and thus cannot be used with the results of Corollary~\ref{cor:unique}.  In fact, due to the jumps in the MTMP, the uniqueness condition of Theorem~\ref{thm:unique} cannot hold globally, though we can still guarantee a maximal (and minimal) clearing solution.
\item If $f(\gamma) = 1 - b \gamma^a$ then Corollary~\ref{cor:unique} is satisfied so long as $a \leq 1$ and $b \in [0,M^{-a})$.  Though, if $a < 1$ then the condition for $\alpha$ results in an empty interval.  This is confirmed by observing the results of Theorem~\ref{thm:unique} to determine that uniqueness holds so long as $a \geq 1$ with
$\alpha \in \frac{1}{\theta_{\min}} \times (\frac{a(M-\gamma_f)-M(1-bM\gamma_f^{a-1})}{(1+a)(M-\gamma_f)-M(1-bM\gamma_f^{a-1})} \; , \; 1)$
for $\gamma_f \in [0,M]$ solving $f(\gamma) = a(1-\frac{\gamma}{M})$.  Note that $\gamma_f$ is unique if $a \geq 1$ but nonexistent if $a < 1$.
\item If $f(\gamma) = (1 - b\gamma)^a$ then Corollary~\ref{cor:unique} is satisfied for any $b < M^{-1}$ with $ab \geq 0$. The risk-weight $\alpha$ is then constrained by $\alpha \in \frac{1}{\theta_{\min}} \times (\frac{abM}{1+abM} \; , \; 1)$.
\item If $f(\gamma) = \exp(-b\gamma)$ then Corollary~\ref{cor:unique} is satisfied for any $b \geq 0$.  The risk-weight $\alpha$ is then constrained by $\alpha \in \frac{1}{\theta_{\min}} \times (\frac{bM}{1+bM} \; , \; 1)$.
\end{enumerate}
Intriguingly, the two generalizations of the linear inverse demand function provide very different results related to uniqueness of the clearing solution.  Further consideration of this discrepancy and deduction of the appropriate shape of the inverse demand function would be an important follow-up study.
\end{example}

We wish to conclude this section by considering the setting of price makers rather than price takers as implicitly is assumed for the prior results. As described below in~\eqref{eq:equilibrium} and~\eqref{eq:equilibrium-constraint}, this price making equilibrium is formulated comparably to the price taking setting of Example~\ref{ex:gamma}\eqref{ex:gamma-equilibrium} with the distinction being that each bank acts with knowledge of how its actions affect the prices for the regulatory constraint. Specifically, rather than fixing the prices as $(q,\bar q)$ in the regulatory constraint $G(q,\bar q)$ of Example~\ref{ex:gamma}\eqref{ex:gamma-equilibrium}, we now consider the case in which banks fully consider the inverse demand functions in place of $(q,\bar q)$ so as to guarantee that their own actions are acceptable to the regulator when accounting for the price impacts; this is encoded mathematically in~\eqref{eq:equilibrium-constraint}.  Notably, and distinct from e.g.\ \cite{bichuch2020repo}, the price maker equilibrium setting can result in equilibria distinct from the price taking setting of Example~\ref{ex:gamma}\eqref{ex:gamma-equilibrium} and vice versa; this is demonstrated in Section~\ref{sec:div} numerically.
\begin{proposition}\label{prop:equilibrium}
Consider the regulatory and balance sheet assumptions from Section~\ref{sec:setting} with continuously differentiable inverse demand function $F$.
Let $\Gamma^* \in [{\bf 0},S]$ such that every bank follows the price making equilibrium liquidation strategy
\begin{align}
\label{eq:equilibrium} \gamma_i^* &= \argmax_{\gamma_i \in \hat G_i(\sum_{j \neq i} \gamma_{j}^*)} u_i(\gamma_i,\sum_{j \neq i} \gamma_{j}^*)\\
\label{eq:equilibrium-constraint} \hat G_i(\gamma_{-i}^*) &= \left\{\gamma_i \in [\vec{0},s_i] \; \left| \; \begin{array}{l}\left(\bar F(\gamma_i + \gamma_{-i}^*) - [I-A\theta_{\min}]F(\gamma_i + \gamma_{-i}^*)\right)^\T\gamma_i \geq \\ \quad \left(h_i - F(\gamma_i + \gamma_{-i}^*)^\T[I-A\theta_{\min}]s_i\right)^+ \wedge \\ \quad\quad \left[\left(\bar F(\gamma_i + \gamma_{-i}^*) - [I-A\theta_{\min}]F(\gamma_i + \gamma_{-i}^*)\right)^\T s_i\right] \end{array}\right.\right\}
\end{align}
with strictly concave utility function.
Let the inverse demand function be such that
\[\gamma_i \in [\vec{0},s_i] \mapsto \bar F(\gamma_i + \gamma_{-i})^\T \gamma_i+F(\gamma_i + \gamma_{-i})^\T [I-A\theta_{\min}] (s_i-\gamma_i)\]
is concave and strictly increasing for any $\gamma_{-i} \in [\vec{0},\sum_{j \neq i}s_j]$ for every $i \in \ncal$.
There exists a Nash equilibrium liquidation solution $\Gamma^* \in [{\bf 0},S]$ and associated clearing prices $q^* = F(\sum_{i \in \ncal} \gamma_i^*)$ and $\bar q^* = \bar F(\sum_{i \in \ncal} \gamma_i^*)$.
\end{proposition}
\begin{remark}
As with Example~\ref{ex:gamma}\eqref{ex:gamma-utility} and~\eqref{ex:gamma-equilibrium}, if the utility function $u_i$ for bank $i$ is decreasing in the liquidations of that bank then the minimal liquidation condition is satisfied in the price making equilibrium for that bank at the clearing prices.
\end{remark}

\section{Stability and sensitivity analysis}\label{sec:sensitivity}
In this section, we perform sensitivity analysis of the equilibrium prices with respect to the system parameters. This is a critical exercise as the exact system parameters are often uncertain and the results depend on how these parameters are calibrated.  As far as the authors are aware, a systematic study on the dependence of clearing prices to system parameters has not been undertaken previously. We characterize the sensitivity analysis as a fixed point problem and prove the existence and uniqueness of the solution to this problem.  In particular, we are interested in studying the sensitivity of the clearing solution to changes in the risk-weights $\alpha_k$, the regulatory threshold $\theta_{\min}$, and the firm shortfall $h_i$.
Sensitivity to risk-weights provides a first-order approximation for the impacts of a credit downgrade on the health of the financial system.
We study the sensitivity with respect to the regulatory threshold, not because it is potentially unknown, but because it allows us to quantify the cost of regulation to the different banks during a crisis.  We elaborate on this point in Section~\ref{sec:cor}.
Finally, in considering sensitivity with respect to the firm shortfall, we are able to quantify a measure of how uncertainty in the shortfall might impact the health of the financial system.  In addition, we propose a methodology to determine when there are incentives for a bailout from a central bank or from other institutions in the financial system.  We elaborate on this point in Section~\ref{sec:bailout}.  This question was also proposed and studied in~\cite{CLY14}.

\subsection{Sensitivity analysis}\label{sec:derivative}
Throughout this section let $\#$ denote an arbitrary parameter of the fire sale model considered in this work.  For example, $\#$ can denote $\alpha_k$ for some asset $k$.  We utilize this general notation as we will see that the sensitivity analysis can be constructed in this general setting with only minor adjustments.

\begin{theorem}\label{thm:sensitivity}
Consider the setting of Theorem~\ref{thm:unique} with differentiable inverse demand function $F$ in which the liquidation function $\Gamma: \D \to [{\bf 0},S]$ is such that $\gamma_i$ is strictly decreasing on $\{(q,\bar q) \in \D \; | \; i \in \lcal(q,\bar q)\}$ for any bank $i$.  Define $\Gamma^*: \D \to [\vec{0},M]$ as $(q,\bar q) \in \D \mapsto \Gamma(q,\bar q)^\T \vec{1}$.  The partial derivatives describing the sensitivity of clearing prices $(q^*,\bar q^*)$ to the parameter $\#$ are defined by:
\begin{align*}
\left(\begin{array}{c} \pdv{q^*}{\#} \\ \pdv{\bar q^*}{\#} \end{array}\right) &= \left[I - \left(\begin{array}{cc} \diag\left[F'(\Gamma^*(q^*,\bar q^*))\right] & {\bf 0} \\ {\bf 0} & \diag\left[\bar F'(\Gamma^*(q^*,\bar q^*))\right] \end{array}\right) \left(\begin{array}{c} J\Gamma^*(q^*,\bar q^*) \\ J\Gamma^*(q^*,\bar q^*) \end{array}\right)\right]^{-1} \\
&\qquad\qquad \times \left(\begin{array}{c} \diag\left[F'(\Gamma^*(q^*,\bar q^*))\right] \\ \diag\left[\bar F'(\Gamma^*(q^*,\bar q^*))\right] \end{array}\right) \pdv{\Gamma^*(q^*,\bar q^*;\#)}{\#}
\end{align*}
where $J\Gamma^*(q^*,\bar q^*)$ denotes the Jacobian of $\Gamma^*$ at $(q^*,\bar q^*)$ and, to simplify notation, $F'(\Gamma^*) := (f_1'(\Gamma_1^*),...,f_m'(\Gamma_m^*))^\T$ and similarly with $\bar F'(\Gamma^*)$.
\end{theorem}

The representation of $\pdv{q^*}{\#}$ provided in Theorem~\ref{thm:sensitivity} follows from implicit differentiation on the clearing mechanism~\eqref{eq:clearing}.  To verify that this representation is well-defined, in Online Appendix~\ref{app:thm:sensitivity}, we prove the invertibility of the matrix on the right-hand side as a Leontief inverse.  This matrix inverse provides the additional gains or losses caused solely by the clearing mechanism due to contagion effects.


One simple application of the sensitivity analysis provided by Theorem~\ref{thm:sensitivity} is in studying the impacts of (marginal) changes in the risk-weights of the various assets under study in this work.  For instance, if an asset is subject to a credit downgrade then its risk-weight will increase.  Though in reality these take discrete values, we are able to analytically describe the impact to the equilibrium prices $(q^*,\bar q^*)$ of a marginal change in the risk-weight $\alpha_k$ of asset $k$.  In particular, a change in the risk-weight for a single asset can cause cross-impacts to all other assets through the fire-sale mechanism as well.  The total impact to market capitalization, due to the drop in MTMPs, from a parallel jump in risk-weights (e.g., from a market downturn causing a system-wide reevaluation of credit ratings) can be quantified as
\[\sum_{k = 1}^m \pdv{M^\T q^*}{\alpha_k} = M^\T \sum_{k = 1}^m \pdv{q^*}{\alpha_k} \leq 0.\]
As a market downturn increases the risk of credit default events, all risk-weights are subject to shifts during exactly the financial crises studied in this work.  As such, this drop in market capitalization can be considered as a stability check on the clearing prices in order to account for these feedback effects, which are otherwise not considered.

\subsection{Cost of regulation}\label{sec:cor}
A particularly interesting application of the sensitivity analysis provided by Theorem~\ref{thm:sensitivity} is in the development of a scheme for computing the cost of regulation incurred by each bank in a stress scenario. This is based on the idea that a tightened regulatory threshold $\theta_{\min}$, during a stress scenario, will result in an increased loss for a bank due to the pro-cyclical nature of this regulatory environment.  Therefore computing the loss incurred for a marginal increase in the regulatory threshold gives a measure of the regulatory cost for a bank. The loss incurred, and hence the cost of regulation, may be quantified in three different ways which we will consider. In the following computations, we wish to provide a reminder that the shortfall $h_i$ of bank $i$ also depends on $\theta_{\min}$, which must also be taken into account.
\begin{itemize}
\item \textit{Cost of regulation on markets:}
Consider the post-fire sale total market capitalization $M^\T q^*$ where $(q^*,\bar q^*)$ denotes the clearing prices.  As the regulatory threshold $\theta_{\min}$ increases, the stressed banks have to liquidate additional assets to satisfy the tighter regulatory environment.  This causes prices to drop and feedback effects to the system.  Ultimately, these effects are also felt by the market as a whole by causing a drop in the market capitalizations.  Mathematically, the losses to the market capitalization caused by this marginal tightening of the regulatory threshold, denoted by $CR$, is provided by
\begin{align*}
CR &:= \pdv{M^\T(\vec{1} - q^*)}{\theta_{\min}} = -M^\T \pdv{q^*}{\theta_{\min}} \geq 0.
\end{align*}
\item \textit{Cost of regulation from realized losses:}
Let us consider the situation where, under the current regulatory regime, a bank has to liquidate a part of its assets. As the threshold $\theta_{\min}$ increases, that bank has to liquidate even more of its assets to satisfy the capital adequacy requirements. We can use the marginal change in losses from implementing the sales from a marginal change in $\theta_{\min}$ to quantify the cost of regulation. Let $(q^*, {\bar q}^*)$ be the clearing prices and let $\gamma_i(q^*,\bar q^*)$ be the liquidation strategy of bank $i$ under the current regulatory environment $\theta_{\min}$. Mathematically, for bank $i$, we represent cost of regulation from realized loss, denoted $CRL_i$, as
\begin{align*}
CRL_i &:= \pdv{(\vec{1}-{\bar q}^*)^\T\gamma_i(q^*,\bar q^*;\theta_{\min})}{\theta_{\min}} \\
&= -\left(\pdv{\bar q^*}{\theta_{\min}}\right)^\T \gamma_i(q^*,\bar q^*) + \left(\vec{1} - \bar q^*\right)^\T \pdv{\gamma_i(q^*,\bar q^*;\theta_{\min})}{\theta_{\min}} \geq 0.
\end{align*}
We note that $\pdv{\bar q^*}{\theta_{\min}}$ can be computed using the results of Theorem~\ref{thm:sensitivity}, whereas $\pdv{\gamma_i(q^*,\bar q^*;\theta_{\min})}{\theta_{\min}}$ depends on liquidation strategy and can be computed explicitly from that construction.
Further, we wish to note that for the situation where banks are not liquidating any assets (i.e., $i \in \scal(q^*,\bar q^*)$), increasing $\theta_{\min}$ by a marginal amount will not result in increased liquidation losses; indeed in such instances $CRL_i$ is equal to $0$.
\item \textit{Cost of regulation from marked-to-market losses:}
As we noted in the discussion of the cost of regulation from realized losses, increasing $\theta_{\min}$ will not result in increased realized liquidation losses for solvent and liquid banks $i \in \scal(q^*,\bar q^*)$. However, increasing $\theta_{\min}$ might cause some other bank to sell additional illiquid assets which would depreciate the price of these assets. As banks hold overlapping portfolios this causes impacts globally to all other banks due to mark-to-market accounting even if that particular bank was not liquidating assets itself; for bank $i$, we represent cost of regulation from marked-to-market losses, denoted $CMI_i$, as
\begin{align*}
CMI_i &:= -\pdv{\left[x_i+\ell_i+(\bar q^*)^\T \gamma_i(q^*,\bar q^*;\theta_{\min})+(q^*)^\T [s_i-\gamma_i(q^*,\bar q^*;\theta_{\min})]-\bar{p}_i\right]}{\theta_{\min}} \\
&= -\left[\left(\pdv{\bar q^*}{\theta_{\min}}\right)^\T \gamma_i(q^*,\bar q^*) + \left(\pdv{q^*}{\theta_{\min}}\right)^\T [s_i - \gamma_i(q^*,\bar q^*)] + \left(\bar q^* - q^*\right)^\T \pdv{\gamma_i(q^*,\bar q^*;\theta_{\min})}{\theta_{\min}}\right] \geq 0.
\end{align*}
Thus for bank $i$, $CMI_i$ is the (negative of the) sensitivity of its equity with respect to $\theta_{\min}$. Hence $CMI_i$ captures the losses that are not reflected in $CRL_i$. We note that in the situation, where \emph{none} of the banks need to liquidate a fraction of their assets (i.e., if $\scal(q^*,\bar q^*) \cup \dcal(q^*,\bar q^*) = \ncal$), increasing $\theta_{\min}$ will not result in any price depreciation or mark-to-market losses. Thus only in that case do we find that $CMI_i=0$ for every bank $i$.
\end{itemize}

\subsection{Value of rescue funds}\label{sec:bailout}
Another interesting application of the sensitivity analysis provided by Theorem~\ref{thm:sensitivity} is in quantifying the value of providing a marginal bailout to the financial system.
We define a ``central bank bailout'' to be an \emph{external} rescue fund used to prop up the health of the financial system, whereas a ``private firm bailout'' has funds made available by other banks in the financial system.  We consider two possible structures for the use of these rescue funds: direct rescue by providing extra capital to a distressed bank or indirect rescue by purchasing troubled assets.
Such strategies were implemented during the 2007-2009 financial crisis, as well as a decade earlier in the bailout of Long-Term Capital Management when 14 financial institutions (under supervision of the Federal Reserve) determined it less costly to absorb the fund's portfolio than to realize the large mark-to-market losses.
We will consider both sources of bailouts along with the appropriate determinations for when these strategies would be appropriate.  Compare these results with, e.g.,~\cite{bernard2017bail} which similarly studies the incentives and decisions related to rescue funds.  As noted above, we also wish to compare these results with~\cite{CLY14} which studies the value of a rescue fund in the setting \emph{without} leverage.

\subsubsection{Direct bailouts}\label{sec:bailout-direct}
By providing a small amount of additional (liquid) assets to a bank, the capital adequacy requirement becomes easier to attain and fewer assets need to be liquidated.  In doing so, the clearing MTMPs will be improved.  Of note, it is possible that providing a marginal amount of additional capital to a solvent but illiquid bank $i \in \lcal(q^*,\bar q^*)$ can have outsized effects on the health of the system via the feedback effects inherent in the clearing mechanism.
\begin{itemize}
\item \textit{Value of a direct central bank bailout of bank $i$:}
By providing additional capital to a bank, that bank improves its balance sheet and therefore need not sell as many assets.  This causes the clearing MTMPs to grow, thus also raising the (time $t = 1$) market capitalization of the various assets.  Of note, it is possible that providing a marginal bailout (to a solvent but illiquid firm $i \in \lcal(q^*,\bar q^*)$) can provide outsized effects on the market capitalization of the system, thus providing incentives to undertake this action.  In particular, if the impacts of the bailout are larger than the initial cost of the bailout, then such incentives exist.  Herein we will measure the value to society of a bailout by the impact on the total market capitalization.
Specifically, as a bailout to bank $i$ increases the liquid holdings $x_i$, the value of a (marginal) direct bailout is the difference between the gains to market capitalization due to reducing shortfall $h_i$ and the costs of the bailout itself.
Mathematically, we define this bailout decision structure as
\begin{align*}
DCB_i &:= -\left(\pdv{M^\T q^*}{h_i} + 1\right) = -\left(M^\T \pdv{q^*}{h_i} + 1\right).
\end{align*}
In particular, the sign of $DCB_i$ is important.  If $DCB_i > 0$ then a (marginal) bailout should be undertaken to prop up bank $i$, otherwise the bailout ultimately costs more than it benefits the system.  Notably, this would only be given to firms that are solvent but illiquid $\lcal(q^*,\bar q^*)$ (but possibly not all of those such banks).
\item \textit{Value of a direct private firm bailout of bank $i$ from bank $j$:}
In contrast to the central bank bailout considered above, a private firm bailout from bank $j \not\in \dcal(q^*,\bar q^*)$ to bank $i$ is a cash payment from $j$ to $i$.  This type of payment simultaneously decreases the assets available to bank $j$ but improves the health of bank $i$.  In particular, to consider such a bailout, the capital of firm $j$ is studied given these positive and negative changes to the shortfall of banks $j$ and $i$ respectively.
Recall that $h_j = \bar p_j - x_j - (1-\alpha_{\ell,j}\theta_{\min})\ell_j$, therefore we can rewrite the capital of bank $j$ in terms of the shortfall $h_j$ by $ C_j = -h_j + (\bar q^*)^\T \gamma_j(q^*,\bar q^*;h_j)+(q^*)^\T [s_j-\gamma_j(q^*,\bar q^*;h_j)]+\alpha_{\ell,j}\theta_{\min}\ell_j$.  As such, similar to the study of central bank bailouts, we can define the private firm bailout decisions via
\begin{align*}
DPB_{ji} &:= \pdv{C_j(h_j)}{h_j} - \pdv{C_j(h_i)}{h_i}\\
\begin{split} &= \left(\pdv{\bar q^*}{h_j} - \pdv{\bar q^*}{h_i}\right)^\T \gamma_j(q^*,\bar q^*) + \left(\pdv{q^*}{h_j} - \pdv{q^*}{h_i}\right)^\T [s_j - \gamma_j(q^*,\bar q^*)]\\
    &\qquad + \left(\bar q^* - q^*\right)^\T \left[\pdv{\gamma_j(q^*,\bar q^*;h_j)}{h_j} - \pdv{\gamma_j(q^*,\bar q^*;h_i)}{h_i}\right] - 1. \end{split}
\end{align*}
We wish to note that, typically, $\pdv{\gamma_j(q^*,\bar q^*;h_i)}{h_i} = 0$ as the behavior of bank $j$ does not depend directly on the shortfall of bank $i$; we refer back to the, e.g., the proportional liquidation strategy of Example~\ref{ex:gamma}\eqref{ex:gamma-proportional}.
As with the central bank bailout above, if $DPB_{ji} > 0$ then bank $j$ has the incentive to provide a marginal rescue fund to bank $i$.  Notably this would only be given to firms that are solvent but illiquid $\lcal(q^*,\bar q^*)$ (but possibly not all of those such banks).
\end{itemize}

\subsubsection{Indirect bailouts}\label{sec:bailout-indirect}
In contrast to the direct bailouts mentioned above, it might be more politically palatable to directly \emph{purchase} distressed assets; when undertaken by a central bank, these indirect bailouts are akin to the actions of a market maker of last resort.  This kind of indirect bailout was undertaken in the United States following the 2008 financial crisis through the Troubled Asset Relief Program.  In particular, by providing additional liquidity to an asset, the health of each individual firm will improve through feedbacks inherent in the clearing mechanism.  We wish to note that, in this setting, we are considering a modified version of the sensitivity analysis presented in Theorem~\ref{thm:sensitivity}.  To consider this setting we need to introduce the modified inverse demand function with the troubled asset purchasing.  Let $\beta \in \bbr^m_+$ be the cash invested in this indirect rescue of asset $k$, then the modified clearing problem can be written as:
\[(q,\bar q) = \left(F\left(\left[\sum_{i \in \ncal} \gamma_i(q,\bar q) - \diag[\bar q]^{-1}\beta\right]^+\right) \; , \; \bar F\left(\left[\sum_{i \in \ncal} \gamma_i(q,\bar q) - \diag[\bar q]^{-1}\beta\right]^+\right)\right);\]
as the rescue fund is also monotonic, the uniqueness argument presented in Theorem~\ref{thm:unique} still holds.  In considering this bailout setting, we are interested in taking the derivative w.r.t.\ the rescue fund $\beta_k$ at $\beta = \vec{0}$.  Implicit differentiation and rearranging terms provides us with the form for $\pdv{q^*}{\beta_k}$ and $\pdv{\bar q^*}{\beta_k}$ as the same as that given in Theorem~\ref{thm:sensitivity}, but with new right-hand side, i.e.,
\begin{align*}
\left(\begin{array}{c} \pdv{q^*}{\beta_k} \\ \pdv{\bar q^*}{\beta_k} \end{array}\right) &= -\left[I - \left(\begin{array}{cc} \diag\left[F'(\Gamma^*(q^*,\bar q^*))\right] & {\bf 0} \\ {\bf 0} & \diag\left[\bar F'(\Gamma^*(q^*,\bar q^*))\right] \end{array}\right) \left(\begin{array}{c} J\Gamma^*(q^*,\bar q^*) \\ J\Gamma^*(q^*,\bar q^*) \end{array}\right)\right]^{-1} \\
&\qquad\qquad \times \left(\begin{array}{c} \frac{f_k'(\Gamma_k^*(q^*,\bar q^*))}{\bar q_k^*}e_k \\ \frac{\bar f_k'(\Gamma_k^*(q^*,\bar q^*))}{\bar q_k^*}e_k \end{array}\right)
\end{align*}
where $e_k \in \bbr^m$ is the unit vector with a single $1$ in its $k^{th}$ component.
\begin{itemize}
\item \textit{Value of an indirect central bank bailout of asset $k$:}
By providing additional liquidity to an asset, the MTMP and VWAP of that asset would improve.  Immediately this causes the clearing prices to grow and, due to the pro-cyclical nature of the capital adequacy ratio, results in banks liquidating fewer assets as well.  Therefore such an indirect bailout can easily have outsized benefits to the health of the financial system, even to assets other than the one getting the additional liquidity; this provides an incentive for this bailout to take place.  As before, if the impacts of the bailout are larger than its initial cost, then the incentive structure is in place for a bailout to occur.  Also as before, herein we will measure the value to society of a bailout by the impact on the total market capitalization.  Mathematically, we define this bailout decision structure via
\begin{align*}
ICB_k &:= \pdv{M^\T q^*}{\beta_k} - 1 = M^\T \pdv{q^*}{\beta_k} - 1.
\end{align*}
In particular, the sign of $ICB_i$ is important.  If $ICB_i > 0$ then a (marginal) bailout should be undertaken to provide extra liquidity to asset $k$, otherwise the bailout ultimately costs more than it benefits the system.  Notably, this would only be given to assets in distress $q_k^* < 1$ (but possibly not all of those such assets).
\item \textit{Value of an indirect private firm bailout of asset $k$ from bank $j$:}
As with the indirect central bank bailout, an indirect private firm bailout provides additional liquidity to a specific asset.  This additional market liquidity can cause large feedback gains on the clearing prices.  In order to participate in such a bailout, we assume that the bank is solvent, i.e., $j \not\in \dcal(q^*,\bar q^*)$.  Such a bank would decide to participate in an indirect bailout if the effects from providing the liquidity (on firm capital) and holding marginal more units of asset $k$ outweigh the costs from also increasing shortfall in tandem.  As such, the value of an indirect bailout from firm $j$ on asset $k$ can be computed mathematically via
\begin{align*}
IPB_{jk} &:= \pdv{C_j(h_j)}{h_j} + \frac{1}{\bar q_k} \pdv{C_j(s_{jk})}{s_{jk}} + \left.\pdv{C_j(\beta_k)}{\beta_k}\right|_{\beta = \vec{0}}\\
\begin{split} &= \left(\pdv{\bar q^*}{h_j} + \frac{1}{\bar q_k^*}\pdv{\bar q^*}{s_{jk}} + \pdv{\bar q^*}{\beta_k}\right)^\T \gamma_j(q^*,\bar q^*) + \left(\pdv{q^*}{h_j} + \frac{1}{\bar q_k^*}\pdv{q^*}{s_{jk}} + \pdv{q^*}{\beta_k}\right)^\T [s_j - \gamma_j(q^*,\bar q^*)] \\
    &\qquad + (\bar q^* - q^*)^\T \left[\pdv{\gamma_j(q^*,\bar q^*;h_j)}{h_j} + \frac{1}{\bar q_k^*} \pdv{\gamma_j(q^*,\bar q^*;s_{jk})}{s_{jk}}\right] - \left(1 - \frac{q_k^*}{\bar q_k^*}\right). \end{split}
\end{align*}
As with the indirect central bank bailout above, if $IPB_{jk} > 0$ then bank $j$ has the incentive to provide a marginal rescue fund to prop up the value of asset $k$.  Notably, this would only be given to assets in distress $q_k^* < 1$ (but possibly not all of those such assets).
\end{itemize}

\begin{remark}\label{rem:indirect}
We conjecture that a direct bailout (of a solvent but illiquid institution) will typically outperform an indirect bailout of the assets being held.  This is due to leveraging effects, i.e., if a bank is given additional capital, through leverage, they can \emph{avoid} (in first order effects) liquidating a greater value of assets than they obtained in capital in the first place.  An indirect bailout will always (in first order effects) compensate for exactly the value of the bailout.  As leverage ratios are typically larger than 1, we make this conjecture.
\end{remark}

\section{Case studies}\label{sec:case}
In this section we consider three case studies to discuss the implications of our  model. For simplicity, each of the case studies is undertaken with a linear inverse demand function. We restrict the risk-weights $\alpha$ to the bound discussed in Corollary~\ref{cor:unique} and Example~\ref{ex:unique-idf}. Briefly, the three case studies are as follows:
\begin{enumerate}
\item First, we consider a simple single asset, two bank system in order to explore some simple implications of liquidations and price impacts on the failure of different institutions.
\item Second, we consider a two asset, two bank system in order to explore the implications of diversification under different liquidation functions.  We compare this case study with that of, e.g.,~\cite{CW19}.
\item Third, we consider a system of six large banks.  This data is loosely calibrated to the 2015 CCAR stress test as considered in~\cite{BW19}. We use this data to study the both the cost of regulation and the value of rescue funds in an illustrative financial system.
\end{enumerate}

\subsection{Fire sale implications on failures}\label{sec:discussion}
In this case study, we wish to consider some properties of the contagion on financial stability.  To do so in a simple setting, we consider a two bank ($n=2$) and single asset ($m=1$) system.  We assume that the banks do not hold any liquid or non-marketable asset, i.e., $x_i=\ell_i=0$ for $i=1,2$.  Additionally, we assume that both banks hold a single unit of the marketable illiquid asset, i.e., $s_i=1$ for $i=1,2$; as there is a single marketable illiquid asset, the liquidation strategy for both banks follows the structure given in Example~\ref{ex:gamma}\eqref{ex:gamma-1asset}.  The balance sheet is completed with liabilities of $\bar p_1 = 0.9$ and $\bar p_2 = 0.6$.
Finally, consider the regulatory environment with $\theta_{\min} = 0.2$ and with risk-weight $\alpha = 1$.

With the initial price $q = \bar q = 1$ at time $t = 0$, the original (shocked) capital adequacy ratios are
\begin{align*}
\theta_1(0) &= 0.1 < \theta_{\min}\\
\theta_2(0) &= 0.4 > \theta_{\min}.
\end{align*}
Notably, even though its initial capital adequacy ratio is below the threshold, $1 \in \lcal(1,1)$ as this bank (at the initial prices of $1$) is able to liquidate a fraction ($\gamma_1(1,1) = 0.5$) of its portfolio in order to stay solvent; in contrast bank $2$ is solvent and liquid at these initial prices.

To consider the implications of the fire sales, we now wish to consider the price impacts.  In particular, we assume that all liquidations cause linear price impacts $F(\Gamma) = 1 - b\Gamma$ for price impact parameter $b \in (0.125,0.5)$ to guarantee uniqueness of the clearing prices as in Corollary~\ref{cor:unique}.  In particular, we will choose \emph{two} price impacts in this range.
\begin{itemize}
\item \emph{Low price impacts $b = 0.15$}: Ansatz take bank $1$ to be solvent but illiquid and bank $2$ to be solvent and liquid.  That is, the clearing solution is such that
    \begin{equation*}
    \left.\begin{array}{rl}
    q^* &= 1 - 0.15\left(\frac{0.9 - 0.8q^*}{\bar q^* - 0.8q^*}\right)\\
    \bar q^* &= 1 - 0.075\left(\frac{0.9 - 0.8q^*}{\bar q^* - 0.8q^*}\right)
    \end{array}\right\} \quad \Rightarrow \quad
    \left\{\begin{array}{rl}
    q^* &= 1 - 0.15\left(\frac{0.9 - 0.8q^*}{0.5 - 0.3q^*}\right)\\
    \bar q^* &= 0.5\left(1+q^*\right)
    \end{array}\right.
    \end{equation*}
    which can be solved analytically.  In particular, only one solution exists with prices less than $1$ given by
    \[q^* = \frac{1}{30}\left(34 - \sqrt{61}\right) \approx 0.8730 \quad \text{ and } \quad \bar q^* = \frac{1}{60}\left(64 - \sqrt{61}\right) \approx 0.9365.\]
    It remains to test the ansatz to make sure that this is truly a clearing solution.  Under these prices $\gamma_1(q^*,\bar q^*) \approx 0.8467$ and $\gamma_2(q^*,\bar q^*) = 0$, which validates our initial guess.  This setting presents a simple clearing problem in which, though bank $1$ is ``insolvent'' at time $t = 0$, through liquidations it is able to become solvent through liquidations.  We wish to note that the minimal liquidation condition guarantees that, generally, any bank with $\theta_i(0) < \theta_{\min}$ will, at the clearing prices, satisfy the inequality $\theta_i(0) \leq \theta_{\min}$ with the chance of recovering solvency as seen in this example.
\item \emph{High price impacts $b = 0.45$}: Ansatz take both banks to be insolvent.  That is, the clearing solution is given by
    \begin{equation*}
    q^* = 0.10 \quad \text{ and } \bar q^* = 0.55.
    \end{equation*}
    Under these prices $\gamma_1(q^*,\bar q^*) = 1$ and $\gamma_2(q^*,\bar q^*) = 1$ which validates our initial guess.  This setting presents a simple clearing problem in which, though bank $2$ is ``solvent and liquid'' at time $t = 0$, through price-mediated contagion it becomes insolvent at time $t = 1$.  This demonstrates that, generally, contagion can cause an otherwise solvent bank to become illiquid or insolvent.
\end{itemize}

\subsection{Effects of liquidation strategies and diversification}\label{sec:div}
In this case study, we consider a two bank ($n=2$) and two asset ($m=2$) system. We assume that the banks do not hold any liquid or non-marketable asset, i.e., $x_i=\ell_i=0$ for $i=1,2$. We assume that both banks have liabilities $\bar p_i=1$ and the total (pre-fire sale) market capitalization of each asset is 2, i.e., $M_k=s_{1k}+s_{2k}=2$ for $k=1,2$.

We study the impact of diversification in this system by varying the composition of the illiquid asset holdings of each bank. To do this, we use a similar setting to Example 5.4 of~\cite{feinstein2019leverage}. That is, consider a parameter $\lambda \in [0,2]$ and set $s_{11}=\lambda$, $s_{12}=M_2-\lambda$, $s_{21}=M_1-\lambda$, and $s_{22}=\lambda$. When $\lambda\in\{0,2\}$, the banks are holding non-overlapping portfolios, corresponding to a \emph{fully diverse system}. When $\lambda=1$, the portfolios of the banks are identical, corresponding to a \emph{fully diversified system}. Due to symmetry between the banks, we will only consider $\lambda \in [0,1]$. Thus as $\lambda$ increases, the system moves from fully diverse to fully diversified. Though the sensitivity of the clearing prices to $\lambda$ can be computed as in Theorem~\ref{thm:sensitivity}, herein we focus on numerical simulations only as we wish to highlight the impacts of liquidation strategies on the effects of diversification in comparison with, e.g.,~\cite{CW19,detering2018suffocating}.

For the purpose of this example, we will consider the linear inverse demand functions $f_k(\Gamma_k) = 1 - \frac{\Gamma_k}{5}$ for both assets $k=1,2$.  Further, we will consider three liquidation functions in this case study in order to determine the impacts such choices have on the clearing prices and market capitalizations; these liquidation functions are:
\begin{itemize}
\item \emph{proportional liquidation} as discussed in Example~\ref{ex:gamma}\eqref{ex:gamma-proportional};
\item \emph{price taking [PT] equilibrium liquidation} as discussed in Example~\ref{ex:gamma}\eqref{ex:gamma-equilibrium} in which both banks are trying to minimize their realized loss, i.e., $u_i(\gamma_i, \gamma^*_{-i})= -\gamma_i^\T (\vec{1}-\bar F (\gamma_i+\gamma^*_{-i}))$.  We note that this is a strictly decreasing and concave utility function on $\gamma_i \in [\vec{0},s_i]$, therefore the minimal liquidation assumption holds.  Further this is a continuous, strictly concave, and diagonally strictly concave function (see, e.g.,~\cite{bichuch2018borrowing} for proof of such), therefore this strategy exists, is unique, and is continuous in the prices $(q,\bar q)$ as discussed when introduced in Example~\ref{ex:gamma}.
\item \emph{price making [PM] equilibrium liquidation} as discussed in Proposition~\ref{prop:equilibrium} in which both banks are trying to minimize their realized loss, i.e., $u_i(\gamma_i,\gamma^*_{-i})= -\gamma_i^\T (\vec{1}-\bar F(\gamma_i+\gamma^*_{-i}))$.  We note that this is a strictly decreasing and strictly concave utility function on $\gamma_i \in [\vec{0},s_i]$, therefore the minimal liquidation assumption holds.
\end{itemize}
We wish to compare the results for proportional liquidation to standard results, e.g., in~\cite{CW19,detering2018suffocating}.  We present it here for the direct comparison to the equilibrium liquidation strategy which, as we demonstrate below, does \emph{not} exhibit the typical tradeoffs between diversity of bank investments with diversification of the initial trading book.
\begin{remark}\label{rem:equilibrium}
Though the price taking equilibrium liquidation strategy has a clearing solution (Proposition~\ref{prop:exist}\eqref{prop:exist-brouwer}), it need not be unique.  We compute the clearing prices by Picard iterations beginning at $(q^0,\bar q^0) = (\vec{1},\vec{1})$, i.e., no impacts.  We wish to note that this method converged to a clearing solution for every choice of parameters tested hinting at a stronger property than proven thus far.

In contrast, given the select parameters herein, the existence of a clearing solution is \emph{not} guaranteed for the price making equilibrium setting.  However, we compute this clearing solution by Picard iterations beginning at $\Gamma^0 = {\bf 0}$, i.e., no liquidations.  As with the price taking equilibrium setting, we wish to note that this method converged to a clearing solution for every choice of parameters tested hinting at a stronger property than proven thus far.
\end{remark}

\begin{figure}[]
\centering
\begin{subfigure}[t]{0.48\linewidth}
\centering
\includegraphics[width=\linewidth]{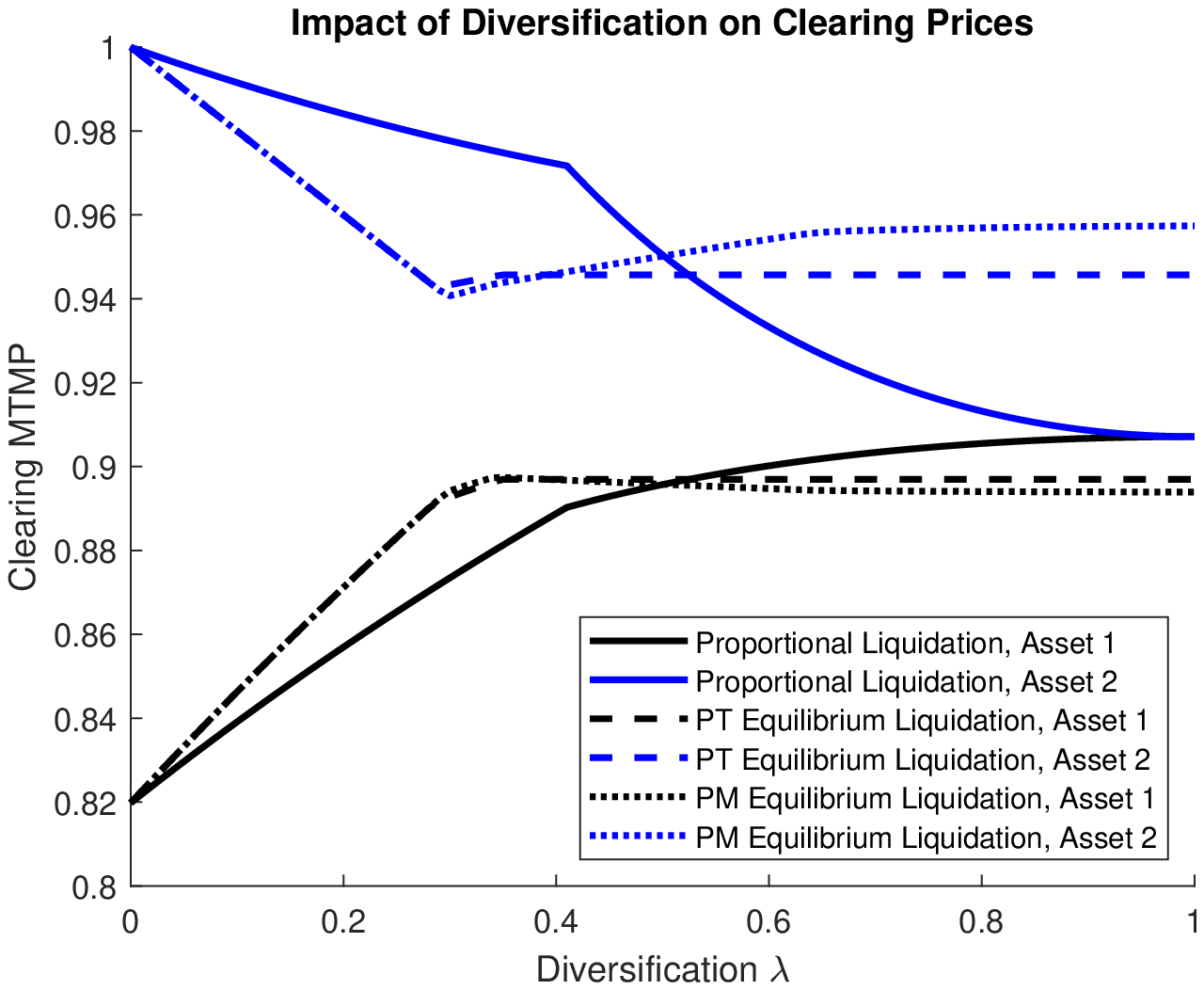}
\caption{Clearing MTMP under proportional and equilibrium liquidation strategies with varying levels of diversification.}
\label{fig:CS2Q}
\end{subfigure}
~
\begin{subfigure}[t]{0.48\linewidth}
\centering
\includegraphics[width=\linewidth]{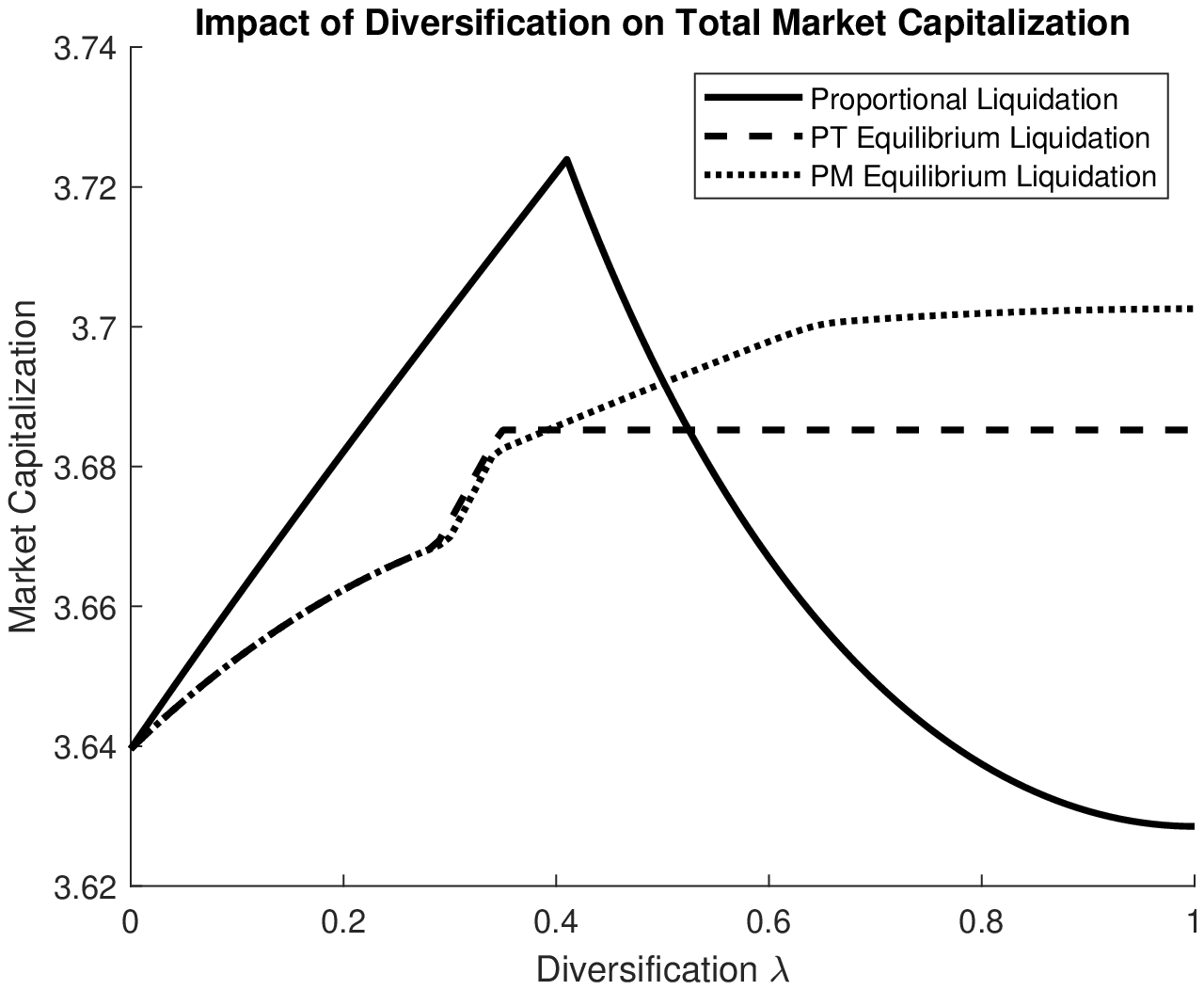}
\caption{Clearing total market capitalization under proportional and equilibrium liquidation strategies with varying levels of diversification.}
\label{fig:CS2MC}
\end{subfigure}
\caption{Section~\ref{sec:div}: The impacts of diversification of portfolio holdings and liquidation strategies on system health.}
\label{fig:CS2}
\end{figure}

Consider the regulatory environment with $\theta_{\min} = 0.2$ and with risk-weights $\alpha_1=\alpha_2 = 2$ at time $t = 0$.  With these parameters, both banks satisfy the capital adequacy requirements at $t = 0$ without any fire sales occurring.  However, consider at time $t = 0^+$ the first asset has a credit downgrade causing its risk-weight to double, i.e., $\alpha_1 = 4$.  This stress precipitates a fire sale of one or both banks depending on their investments at time $t = 1$ in order to satisfy the capital adequacy requirements.  Notably, we consider the stress to the system to be a credit downgrade rather than a shock to the balance sheet of a bank as is typically assumed (see, e.g., \cite{BW19,feinstein2019leverage}).  The results of this fire sale are displayed in Figure~\ref{fig:CS2}, which demonstrate the significant impacts that the choice of liquidation function has on the clearing prices.

First, in Figure~\ref{fig:CS2Q}, we see that \emph{cross-asset contagion} is a significant factor in the proportional liquidation scenario.  That is, even though asset $2$ is not shocked, under proportional liquidation its price monotonically decreases in $\lambda \in [0,1]$.  Both equilibrium liquidation strategies appear to have cross-asset contagion, but only up to a point.  This form of contagion is stronger than the proportional strategy for highly diverse portfolios ($\lambda \leq 0.3$), but reaches a limiting amount of contagion for more diversified portfolios.
Second, in Figure~\ref{fig:CS2MC}, we see that the total market capitalization of the system reaches its maximum, in the proportional liquidation setting, at $\lambda \approx 0.4$.  In fact, the worst case in that setting is for the fully diversified portfolio.  In contrast, both equilibrium liquidation strategies have nondecreasing total market capitalization as diversification increases.  That is, diversification has the stabilizing effects typically assigned to it.  In aggregate, the proportional liquidation strategy outperforms the equilibrium strategy for diverse systems ($\lambda \leq 0.55$ for price taking equilibrium and $\lambda \leq 0.50$ for price making equilibrium) but underperforms for diversified investments ($\lambda \geq 0.55$ for price taking equilibrium and $\lambda \geq 0.50$ for price making equilibrium).
Previous research, which generally consider full diversification harmful from a contagion perspective, may have suffered from the bias of focusing on the proportional liquidation strategy. In actuality, policy should be designed considering the strategic aspect of more realistic liquidation functions.
Finally, we wish to compare the two equilibrium liquidation strategies.  Both these equilibrium strategies take similar forms, but the price making equilibrium outperforms the price taking equilibrium liquidation strategy for diversified investments ($\lambda \geq 0.39$) in total market capitalization.  This is likely due to the freedom of the banks to more completely endogenize their own actions on price impacts.

\subsection{Six bank system}\label{sec:large}
In this case study, we use our model to study a stress test of a six bank financial system.  For this, we use the Comprehensive Capital Analysis and Review (CCAR) 2015 data and consider the six global systemically important banks with large trading operations, i.e., Bank of America, Citigroup, Goldman Sachs, JP Morgan Chase, Morgan Stanley, and Wells Fargo, as was done in~\cite{BW19}. The data for these organizations is shown in Table~\ref{table:CCAR} which has been replicated from~\cite{BW19}.  This data allows us to directly characterize the (pre-fire sale) value of the banks' balance sheets (with liabilities equal to the difference between total assets and capital).
For a detailed discussion of the CCAR dataset we refer to \cite{BW19}.  Though we are calibrating the financial system to a real dataset, the granular data necessary for this analysis is not known to us and, as such, this example is for \emph{illustrative purposes only}.

\begin{table}[t]
{\footnotesize
\begin{tabular}{l *{7}{S[table-format=-1.2]}}
\toprule
 & & \multicolumn{3}{c}{\textbf{Assets}} & \multicolumn{2}{c}{\textbf{Risk-Weighted Assets}} \\
\cmidrule(lr){3-5} \cmidrule(lr){6-7}
 & \textbf{Capital} & \textbf{Liquid} & \textbf{Marketable} & \textbf{Nonmarketable} & \textbf{Marketable} & \textbf{Nonmarketable} \\
\textbf{Bank} & {$C$} & {$x$}  & {$\vec{1}^\T s$}  & {$\ell$} & {$\vec{1}^\T A s$} & {$\alpha_{\ell} \ell$} \\
\midrule
Bank of America & 161.62 & 138.63 & 565.20 & 1400.70 & 279.40 & 1185.60 \\
Citigroup       & 165.45 & 32.11  & 596.90 & 1213.17 & 203.50 & 1089.10 \\
Goldman Sachs   & 90.98  & 57.58  & 473.97 & 324.69  & 335.91 & 234.50  \\
JP Morgan Chase & 206.59 & 26.97  & 857.40 & 1687.90 & 313.40 & 1305.60 \\
Morgan Stanley  & 74.97  & 21.39  & 430.72 & 349.40  & 204.04 & 251.98  \\
Wells Fargo     & 192.90 & 19.60  & 355.95 & 1311.61 & 130.24 & 1115.26 \\ \bottomrule
\end{tabular}
}
\caption{Section~\ref{sec:large}: Assets (in billion of dollars) for the six banks under consideration.}
\label{table:CCAR}
\end{table}
\begin{table}[H]
\begin{tabular}{@{}ccccccccccccccccc@{}}
\toprule
\textbf{Asset} & 1    & 2    & 3   & 4    & 5    & 6    & 7   & 8    & 9    & 10  & 11  & 12   & 13 & 14  & 15   & 16  \\ \midrule
$\alpha$ & 0.07 & 0.08 & 0.1 & 0.12 & 0.15 & 0.18 & 0.2 & 0.25 & 0.35 & 0.5 & 0.6 & 0.75 & 1  & 2.5 & 4.25 & 6.5 \\ \bottomrule
\end{tabular}
\caption{Section~\ref{sec:large}: Risk-weights $\alpha$ for the marketable illiquid assets.}
\label{table:CCAR-RW}
\end{table}

From the CCAR data set we can immediately find the risk-weight $\alpha_{\ell,i}$ of each firm's non-marketable portfolio by dividing the non-marketable risk-weighted assets by the value of the non-marketable assets ($\alpha_{\ell,i}\ell_i / \ell_i$).  In order to calibrate each bank's marketable portfolio $s$, we make use of the risk-weights for commonly traded assets. We assume that there are $m=16$ illiquid marketable asset and choose $\alpha$ for these $16$ assets to cover a wide range of risk-weights as depicted in Table~\ref{table:CCAR-RW}.
For each bank $i$, the individual portfolio $s_i$ is chosen as the minimizer of the following minimum norm problem
\[\min_{\tilde s_i \in \bbr^m_+} \left\{\|\tilde s_i\|_2 \; \left| \; \vec{1}^\T\tilde s_i = \vec{1}^\T s_i, \; \vec{1}^\T A \tilde s_i = \vec{1}^\T A s_i \right.\right\}\]
where the value of the assets $\vec{1}^\T s_i$ and the risk-weighted assets $\vec{1}^\T A \tilde s_i$ are provided in Table~\ref{table:CCAR}.
Additionally, we use these risk-weights to calibrate the \emph{linear} inverse demand functions $f_k(\Gamma_k) = 1 - b_k \Gamma_k$ by setting $b_k = \frac{4\alpha_k\theta_{\min}}{5(1-\alpha_k\theta_{\min})M_k}$; this choice of liquidity parameter guarantees the uniqueness properties of Corollary~\ref{cor:unique}.
Finally, in accordance with Basel II regulations, we set $\theta_{\min} = 0.08$.
We wish to emphasize that the purpose of this calibration is to provide a demonstrative data set for the case study. An accurate calibration of the financial system is an interesting problem in itself and beyond the scope of the current work.

For simplicity and comparison to prior literature (e.g.,~\cite{GLT15}), we assume all banks follow a proportional liquidation strategy.
Under the setting considered, no fire sale occurs.  This validates our modeling assumptions as no large market event occurred during the period this data covers.
Consider now a $5\%$ shock to the non-marketable assets $\ell_i$ for each of the 6 banks, i.e., the realized non-marketable assets are $95\%$ of the value reported in Table~\ref{table:CCAR}. Under this stress regime, four banks (Citigroup, Goldman Sachs, Morgan Stanley and Wells Fargo) do not need to liquidate any assets in the fire sale.  However, JP Morgan Chase is solvent but illiquid, and Bank of America is insolvent in this stress scenario.

\begin{figure}[H]
\centering
\includegraphics[width=.6\linewidth]{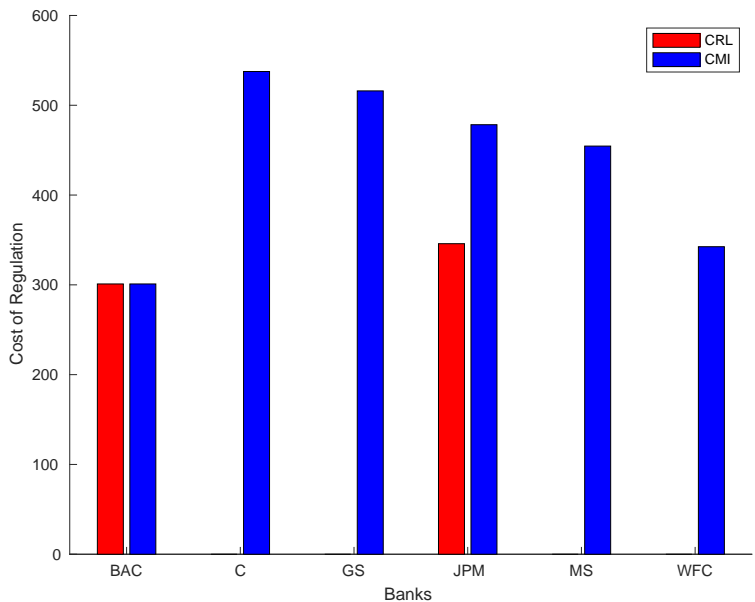}
\caption{Section~\ref{sec:large}: Cost of Regulation $CRL$ and $CMI$.}
\label{fig:CS3}
\end{figure}

With this stress scenario, we first wish to provide the costs of regulation (as discussed in Section~\ref{sec:cor}).  This is displayed in Figure~\ref{fig:CS3}.  Consistent with the theory, only Bank of America and JP Morgan Chase show a non-zero $CRL$, all 6 banks have a non-zero $CMI$. This highlights that even though four banks are not liquidating under the current stress regime, they will incur mark-to-market losses if $\theta_{\min}$ is increased.  We also wish to highlight the costs incurred to the market as a whole due to increased regulatory oversight $CR = 3276.8$ far exceeds the costs incurred by any individual bank.

Finally, we wish to consider the value of direct and indirect rescues of this system.  For direct rescues, we look only at the values of supporting JP Morgan Chase, as all other banks are either liquid or insolvent and, as such, would be unaffected by a marginal influx of liquidity.  Notably, we find that there is a benefit to society for providing such a central bank bailout to JP Morgan Chase with $DCB_{JPM} = 0.3507$.  However, no solvent and liquid institution benefits enough from joining a private firm bailout in this scenario ($DPB_{\cdot,JPM} \approx -0.8$).
As conjectured in Remark~\ref{rem:indirect}, the value of the indirect central bank bailout of every asset is below that of the direct central bank bailout of JP Morgan Chase.  In fact, the value of the indirect central bank bailout is negative (with the high risk-weighted assets having $ICB$ closer to 0 but still negative).  We did not compute the indirect private firm bailout values due to the combination of results for the direct private firm bailouts and indirect central bank bailouts.

\section{Conclusion}\label{sec:conclusion}
In this work we proposed a model for price-mediated contagion due to risk-weight based capital adequacy requirements.  In order to more accurately construct this model, we introduced a two-tier pricing structure: the volume weighted average price (obtained during liquidation of assets) and the mark-to-market price (to account for all unsold assets).  By introducing this pricing structure, we proved the existence and uniqueness of clearing prices; uniqueness had never been proven previously in a leverage of capital ratio requirement based setting of price-mediated contagion.  Through the use of the uniqueness criteria, we provided conditions on the risk-weights based on the liquidity of the assets; this is an improvement over the current, heuristic, method for determining these risk-weights.  Additionally, under the conditions of uniqueness, we determined the sensitivity of the clearing prices to the system parameters.  This sensitivity is valuable due to the uncertainty of many system parameters.  We highlighted two uses for the sensitivity analysis: the cost of regulation during a crisis and the value of a rescue fund.  These results are utilized in two case studies.  The first considers the typical study of the impact of diversification on financial stability; in contrast to prior studies, we demonstrated that the usual result on the tradeoff of diversification with diversity of investment strategies no longer holds when an optimization based liquidation strategy is considered.  We concluded this work with a simulation of a financial system loosely calibrated to the CCAR 2015 stress test data and considered, for that system, the cost of regulation during a stress scenario as well as the value of rescue funds.

\bibliographystyle{plainnat}
\bibliography{bibtex2}

\begin{thebibliography}{27}
\providecommand{\natexlab}[1]{#1}
\providecommand{\url}[1]{\texttt{#1}}
\expandafter\ifx\csname urlstyle\endcsname\relax
  \providecommand{\doi}[1]{doi: #1}\else
  \providecommand{\doi}{doi: \begingroup \urlstyle{rm}\Url}\fi

\bibitem[Acerbi and Scandolo(2008)]{AS08}
Carlo Acerbi and Giacomo Scandolo.
\newblock Liquidity risk theory and coherent measures of risk.
\newblock \emph{Quantitative Finance}, 8\penalty0 (7):\penalty0 681--692, 2008.

\bibitem[Adrian and Shin(2010)]{AS10}
Tobias Adrian and Huyn Shin.
\newblock Liquidity and leverage.
\newblock \emph{Federal Reserve Bank of New York Staff Reports}, 2010.

\bibitem[Aliprantis and Border(2007)]{AB07}
Charalambos~D. Aliprantis and Kim~C. Border.
\newblock \emph{Infinite Dimensional Analysis: A Hitchhiker's Guide}.
\newblock Springer, 2007.
\newblock ISBN 9783540326960.
\newblock URL \url{http://books.google.com/books?id=4hIq6ExH7NoC}.

\bibitem[Almgren and Chriss(2001)]{almgrenchriss2001}
Robert Almgren and Neil Chriss.
\newblock Optimal execution of portfolio transactions.
\newblock \emph{Journal of Risk}, 3:\penalty0 5--40, 2001.

\bibitem[Amini et~al.(2016)Amini, Filipovi\'{c}, and Minca]{AFM16}
Hamed Amini, Damir Filipovi\'{c}, and Andreea Minca.
\newblock Uniqueness of equilibrium in a payment system with liquidation costs.
\newblock \emph{Operations Research Letters}, 44\penalty0 (1):\penalty0 1--5,
  2016.

\bibitem[Bernard et~al.(2017)Bernard, Capponi, and Stiglitz]{bernard2017bail}
Benjamin Bernard, Agostino Capponi, and Joseph~E Stiglitz.
\newblock Bail-ins and bail-outs: Incentives, connectivity, and systemic
  stability.
\newblock Working Paper 23747, National Bureau of Economic Research, 2017.

\bibitem[Bichuch and Feinstein(2019)]{bichuch2018borrowing}
Maxim Bichuch and Zachary Feinstein.
\newblock Optimization of fire sales and borrowing in systemic risk.
\newblock \emph{SIAM Journal on Financial Mathematics}, 10\penalty0
  (1):\penalty0 68--88, 2019.

\bibitem[Bichuch and Feinstein(2020)]{bichuch2020repo}
Maxim Bichuch and Zachary Feinstein.
\newblock A repo model of fire sales with {VWAP} and {LOB} pricing mechanisms.
\newblock 2020.
\newblock Working paper.

\bibitem[Braouezec and Wagalath(2019)]{BW19}
Yann Braouezec and Lakshithe Wagalath.
\newblock Strategic fire-sales and price-mediated contagion in the banking
  system.
\newblock \emph{European Journal of Operational Research}, 274\penalty0
  (3):\penalty0 1180--1197, 2019.

\bibitem[Capponi and Larsson(2015)]{CL15}
Agostino Capponi and Martin Larsson.
\newblock Price contagion through balance sheet linkages.
\newblock \emph{Review of Asset Pricing Studies}, 5\penalty0 (2):\penalty0
  227--253, 2015.

\bibitem[Capponi and Weber(2020)]{CW19}
Agostino Capponi and Marko Weber.
\newblock Systemic portfolio diversification.
\newblock 2020.
\newblock Working paper.

\bibitem[Chen et~al.(2016)Chen, Liu, and Yao]{CLY14}
Nan Chen, Xin Liu, and David~D. Yao.
\newblock An optimization view of financial systemic risk modeling: The network
  effect and the market liquidity effect.
\newblock \emph{Operations Research}, 64\penalty0 (5), 2016.

\bibitem[Cifuentes et~al.(2005)Cifuentes, Shin, and Ferrucci]{CFS05}
Rodrigo Cifuentes, Hyun~Song Shin, and Gianluigi Ferrucci.
\newblock Liquidity risk and contagion.
\newblock \emph{Journal of the European Economic Association}, 3\penalty0
  (2-3):\penalty0 556--566, 2005.

\bibitem[Cont and Schaanning(2017)]{CS17}
Rama Cont and Eric Schaanning.
\newblock Fire sales, indirect contagion and systemic stress testing.
\newblock 2017.
\newblock Norges Bank Working paper.

\bibitem[Cont and Schaanning(2019)]{CS19}
Rama Cont and Eric Schaanning.
\newblock Monitoring indirect contagion.
\newblock \emph{Journal of Banking and Finance}, 104:\penalty0 85--102, 2019.

\bibitem[Detering et~al.(2020)Detering, Meyer-Brandis, Panagiotou, and
  Ritter]{detering2018suffocating}
Nils Detering, Thilo Meyer-Brandis, Konstantinos Panagiotou, and Daniel Ritter.
\newblock Suffocating fire sales.
\newblock 2020.
\newblock Working paper.

\bibitem[Diamond and Rajan(2011)]{DR11}
Douglas Diamond and Raghuram Rajan.
\newblock Fear of fire sales, illiquidity seeking, and credit freezes.
\newblock \emph{The Quarterly Journal of Economics}, 126\penalty0 (2):\penalty0
  557--591, 2011.

\bibitem[Duarte and Eisenbach(2018)]{DE18}
Fernando Duarte and Thomas Eisenbach.
\newblock Fire-sale spillovers and systemic risk.
\newblock \emph{Federal Reserve Bank of New York Staff Reports}, 2018.

\bibitem[Feinstein(2017)]{feinstein2015illiquid}
Zachary Feinstein.
\newblock Financial contagion and asset liquidation strategies.
\newblock \emph{Operations Research Letters}, 45\penalty0 (2):\penalty0
  109--114, 2017.

\bibitem[Feinstein(2020)]{feinstein2019leverage}
Zachary Feinstein.
\newblock Capital regulation under price impacts and dynamic financial
  contagion.
\newblock \emph{European Journal of Operational Research}, 281\penalty0
  (2):\penalty0 449--463, 2020.

\bibitem[Feinstein and El-Masri(2017)]{feinstein2016leverage}
Zachary Feinstein and Fatena El-Masri.
\newblock The effects of leverage requirements and fire sales on financial
  contagion via asset liquidation strategies in financial networks.
\newblock \emph{Statistics and Risk Modeling}, 34\penalty0 (3-4):\penalty0
  113--139, 2017.

\bibitem[Feinstein and Ha{\l}aj(2020)]{feinstein2020interbank}
Zachary Feinstein and Grzegorz Ha{\l}aj.
\newblock Interbank asset-liability networks with fire sale management.
\newblock 2020.
\newblock Working paper.

\bibitem[Greenwood et~al.(2015)Greenwood, Landier, and Thesmar]{GLT15}
Robin Greenwood, Augustin Landier, and David Thesmar.
\newblock Vulnerable banks.
\newblock \emph{Journal of Financial Economics}, 115\penalty0 (3):\penalty0
  3--28, 2015.

\bibitem[Pohl et~al.(2017)Pohl, Ristig, Schachermayer, and
  Tangpi]{PRST2018dimensional}
Mathias Pohl, Alexander Ristig, Walter Schachermayer, and Ludovic Tangpi.
\newblock The amazing power of dimensional analysis: Quantifying market impact.
\newblock \emph{Market Microstructure and Liquidity}, 3\penalty0
  (03n04):\penalty0 1850004, 2017.
\newblock \doi{10.1142/S2382626618500041}.

\bibitem[Rosen(1965)]{rosen1965}
J.B. Rosen.
\newblock Existence and uniqueness of equilibrium points for concave n-person
  games.
\newblock \emph{Econometrica}, 33\penalty0 (3):\penalty0 520--534, 1965.

\bibitem[ten Raa(2006)]{tenraa_2006}
Thijs ten Raa.
\newblock \emph{The Economics of Input-Output Analysis}.
\newblock Cambridge University Press, 2006.

\bibitem[Weber and Weske(2017)]{AW_15}
Stefan Weber and Kerstin Weske.
\newblock The joint impact of bankruptcy costs, fire sales and cross-holdings
  on systemic risk in financial networks.
\newblock \emph{Probability, Uncertainty and Quantitative Risk}, 2\penalty0
  (1):\penalty0 9, 2017.

\end{thebibliography}
\newpage

\appendix
\section{Proofs from Section~\ref{sec:clearing}}
\subsection{Proof of Proposition~\ref{prop:exist}}
\begin{proof}
\begin{enumerate}
\item This is a trivial application of Brouwer's fixed point theorem.
\item This is a trivial application of Tarski's fixed point theorem.
\end{enumerate}
\end{proof}

\subsection{Proof of Theorem~\ref{thm:unique}}
\begin{proof}
As discussed in the preceding section, a bank $i$ can belong to any of the following three mutually exclusive and exhaustive sets:
    \begin{itemize}
    \item solvent and liquid: $\scal(q, \bar q)=\{i \in \ncal \; | \; h_i \leq q^\T [I-A\theta_{\min}] s_i\}$;
    \item solvent but illiquid: $\lcal(q, \bar q)=\{i \in \ncal \; | \; q^\T [I-A\theta_{\min}] s_i  < h_i < \bar q^\T s_i\}$; or
    \item insolvent: $\dcal(q, \bar q)=\{i \in \ncal \; | \; h_i \geq \bar q^\T s_i\}$.
    \end{itemize}
Using the minimal liquidation condition \eqref{eq:mlc}, under any MTMP and VWAP prices $(q, \bar q) \in \D$ for any bank $i$:
\begin{equation*}
(\bar{q} - [I - A\theta_{\min}]q)^\T \gamma_i(q, \bar q) =
\begin{cases}
0 &\text{if } i \in \scal(q, \bar q)\\
h_i - q^\T [I - A\theta_{\min}] s_i &\text{if } i \in \lcal(q, \bar q)\\
(\bar q - [I - A\theta_{\min}]q)^\T s_i &\text{if } i \in \dcal(q, \bar q).
\end{cases}
\end{equation*}

Using Proposition~\ref{prop:exist}\eqref{prop:exist-tarski}, there exists a greatest and least clearing price $(q^\uparrow, \bar q^\uparrow) \geq (q^\downarrow, \bar q^\downarrow)$. Further from $\Gamma$ nonincreasing, $\Gamma^\uparrow:=\Gamma(q^\uparrow, \bar q^\uparrow))^\T \vec{1} \leq  \Gamma(q^\downarrow, \bar q^\downarrow))^\T \vec{1} =: \Gamma^\downarrow$.

Assume that there does not exist a unique clearing price, i.e., there exists some asset $k$ such that either $q_k^\uparrow > q_k^\downarrow$ or $\bar q_k^\uparrow > \bar q_k^\downarrow$ (and thus $\Gamma^\uparrow_k < \Gamma^\downarrow_k$).  Thus considering that both $(q^\uparrow,\bar q^\uparrow)$ and $(q^\downarrow,\bar q^\downarrow)$ are clearing solutions, we utilize the additional property on the inverse demand functions to find a contradiction:
\begin{align*}
    0 &> [(\bar{q}^\uparrow)^\T \Gamma^\uparrow +(q^\uparrow)^\T[I-A\theta_{\min}](M- \Gamma^\uparrow)]-[(\bar{q}^\downarrow)^\T \Gamma^\downarrow +(q^\downarrow)^\T[I-A\theta_{\min}](M- \Gamma^\downarrow)]\\
    & \geq [(\bar{q}^\uparrow)^\T \Gamma^\uparrow +(q^\uparrow)^\T[I-A\theta_{\min}](\sum_{i=1}^{n} s_i- \Gamma^\uparrow)]-[(\bar{q}^\downarrow)^\T \Gamma^\downarrow +(q^\downarrow)^\T[I-A\theta_{\min}](\sum_{i=1}^{n} s_i- \Gamma^\downarrow)]\\
    &= \sum_{i \in \dcal^\uparrow \cap \dcal^\downarrow}({\bar{q}^\uparrow}-{\bar{q}^\downarrow})^\T s_i + \sum_{i \in \lcal^\uparrow \cap \dcal^\downarrow}(h_i-(\bar{q}^\downarrow)^\T s_i)
    +\sum_{i \in \scal^\uparrow \cap \dcal^\downarrow}([I-A\theta_{\min}]q^\uparrow-\bar q^\downarrow)^\T s_i\\
    &\quad+ \sum_{i \in \lcal^\uparrow \cap \lcal^\downarrow}(h_i-h_i)
    +\sum_{i \in \scal^\uparrow \cap \lcal^\downarrow}((\bar{q}^\uparrow)^\T[I-A\theta_{\min}]s_i - h_i)
    +\sum_{i \in \scal^\uparrow \cap \scal^\downarrow}(q^\uparrow-q^\downarrow)^\T[I-A\theta_{\min}]s_i\\
    &>0
\end{align*}
where $\scal^\uparrow := \scal(q^\uparrow,\bar q^\uparrow)$ and $\scal^\downarrow, \lcal^\uparrow, \lcal^\downarrow, \dcal^\uparrow, \dcal^\downarrow$ are defined likewise.
\end{proof}

\subsection{Proof of Corollary~\ref{cor:unique}}
\begin{proof}
By Theorem~\ref{thm:unique}, we need only prove that $\Gamma^* \in [\vec{0},M] \mapsto \bar F(\Gamma^*)^\T \Gamma^* + F(\Gamma^*)^\T [I - A\theta_{\min}] (M - \Gamma^*)$ is strictly increasing.
By taking derivatives and rearranging terms, this is true if for every asset $k$
\begin{equation}\label{eq:riskweight}
\alpha_k > -\frac{1}{\theta_{\min}}\frac{(M_k-\Gamma_k^*)f'_k(\Gamma_k^*)}{f_k(\Gamma_k^*)-(M_k-\Gamma_k^*)f'_k(\Gamma_k^*)} \quad \forall \; \Gamma_k^* \in [0,M_k].
\end{equation}
The additional condition on the inverse demand function $F$ imposed in this corollary is sufficient to ensure that the right-hand side of \eqref{eq:riskweight}, i.e., $-\frac{1}{\theta_{\min}}\frac{(M_k-\Gamma_k^*)f_k'(\Gamma_k^*)}{f_k(\Gamma_k^*)-(M_k-\Gamma_k^*)f_k'(\Gamma_k^*)}$ is nonincreasing in $\Gamma_k^* \in [0,M_k]$.
 
Thus to ensure \eqref{eq:riskweight}, we require $\alpha_k$ to satisfy the inequality at $\Gamma_k^*=0$. Using this fact and Assumption~\ref{ass:alphatheta}, uniqueness is ensured by the conditions of this corollary.
\end{proof}

\subsection{Proof of Proposition~\ref{prop:equilibrium}}
\begin{proof}
To prove the existence of a price making equilibrium liquidation strategy $\Gamma^* \in [{\bf 0},S]$, consider the game~\eqref{eq:equilibrium} for bank $i$ parameterized by the (aggregate) actions of all other firms $\gamma_{-i}^* \in [\vec{0},\sum_{j \neq i} s_j]$.  By the concavity assumptions, if these optimal liquidations are continuous w.r.t.\ the actions of all other firms then existence follows by an application of Brouwer's fixed point theorem.  We will prove continuity by the Berge maximum theorem for bank $i$'s optimization problem; in particular, this result follows so long as $\hat G_i$ defined in~\eqref{eq:equilibrium-constraint} is set-valued continuous. We will break this proof into two pieces:
\begin{enumerate}
\item $\hat G_i$ is upper continuous by an application of the closed graph theorem (see, e.g., Theorem 17.11 of~\cite{AB07}) due to continuity of the inverse demand functions and compactness of the codomain.
\item To prove that $\hat G_i$ is lower continuous, take $\gamma_i \in \hat G_i(\gamma_{-i})$ for some $\gamma_{-i} \in [\vec{0},\sum_{j \neq i} s_j]$ and consider $\gamma_{-i}^k \in [\vec{0},\sum_{j \neq i} s_j] \to \gamma_{-i}$; we wish to show that there exists some $\gamma_i^k \in \hat G_i(\gamma_{-i}^k)$ for every $k$ such that $\lim_{k \to \infty} \gamma_i^k = \gamma_i$.
    First, if $\gamma_i = s_i$ then take $\gamma_i^k = s_i$ for every $k$ as well; by the assumed monotonicity this is always feasible and therefore $\gamma_i^k \in \hat G_i(\gamma_{-i}^k)$.
    Second, assume $\gamma_i \neq s_i$.  This condition implies that
    \[\bar F(\gamma_i + \gamma_{-i})^\T \gamma_i + F(\gamma_i + \gamma_{-i})^\T[I - A\theta_{\min}](s_i - \gamma_i) \geq h_i.\]
    If this inequality is strict then continuity of the inverse demand functions prove the results.  Therefore we need to consider the case in which this constraint is binding. To simplify this proof, consider a parameterization $\gamma_i(t) := (1-t)\gamma_i + t s_i$.  With this representation, we will seek the representation $\gamma_i^k := \gamma_i\left(g(\gamma_{-i}^k)^+ \wedge 1\right)$ for some function $g: [\vec{0},\sum_{j \neq i} s_j] \to \bbr$.  By the differentiability and strict monotonicity assumptions, the implicit function theorem guarantees the continuity of such a function $g$ along the binding constraint (in a neighborhood around $\gamma_{-i}$) and thus the proof is complete.
\end{enumerate}
\end{proof}

\section{Proofs from Section~\ref{sec:sensitivity}}
\subsection{Proof of Theorem~\ref{thm:sensitivity}}\label{app:thm:sensitivity}
\begin{proof}
Consider the clearing procedure~\eqref{eq:clearing}.  Implicit differentiation w.r.t.\ $\#$ at the equilibrium prices $(q^*,\bar q^*)$ provides the pair of \emph{linear} equations for every asset $k$:
\begin{align*}
\pdv{q_k^*}{\#} &= \pdv{f_k(\Gamma_k^*(q^*,\bar q^*;\#))}{\#}\\
    &= f_k'(\Gamma_k^*(q^*,\bar q^*;\#)) \left[\sum_{l = 1}^m \left(\pdv{\Gamma_k^*(q^*,\bar q^*;\#)}{q_l^*} \pdv{q_l^*}{\#} + \pdv{\Gamma_k^*(q^*,\bar q^*;\#)}{\bar q_l^*} \pdv{\bar q_l^*}{\#}\right) + \pdv{\Gamma_k^*(q^*,\bar q^*;\#)}{\#}\right]\\
\pdv{\bar q_k^*}{\#} &= \pdv{\bar f_k(\Gamma_k^*(q^*,\bar q^*;\#))}{\#}\\
    &= \bar f_k'(\Gamma_k^*(q^*,\bar q^*;\#)) \left[\sum_{l = 1}^m \left(\pdv{\Gamma_k^*(q^*,\bar q^*;\#)}{q_l^*} \pdv{q_l^*}{\#} + \pdv{\Gamma_k^*(q^*,\bar q^*;\#)}{\bar q_l^*} \pdv{\bar q_l^*}{\#}\right) + \pdv{\Gamma_k^*(q^*,\bar q^*;\#)}{\#}\right].
\end{align*}
In matrix notation, this problem reduces to solving the linear system:
\begin{align*}
\left[I - W \right] \left(\begin{array}{c} \pdv{q^*}{\#} \\ \pdv{\bar q^*}{\#} \end{array}\right) &= \left(\begin{array}{c} \diag\left[F'(\Gamma^*(q^*,\bar q^*))\right] \\ \diag\left[\bar F'(\Gamma^*(q^*,\bar q^*))\right] \end{array}\right) \pdv{\Gamma^*(q^*,\bar q^*)}{\#} \\
W &= \left(\begin{array}{cc} \diag\left[F'(\Gamma^*(q^*,\bar q^*))\right] & {\bf 0} \\ {\bf 0} & \diag\left[\bar F'(\Gamma^*(q^*,\bar q^*))\right] \end{array}\right) \left(\begin{array}{c} J\Gamma^*(q^*,\bar q^*) \\ J\Gamma^*(q^*,\bar q^*) \end{array}\right).
\end{align*}
Thus the result is proven so long as $I - W$ is invertible.  Note that $W$ is independent of the choice of parameter $\#$.

We will now prove that $W$ is an invertible matrix by considering standard input-output analysis.  To do so, we first wish to point out that $W \geq {\bf 0}$ as it is defined as the multiplication of two non-positive matrices.

Recall that bank $i$ can belong to any of the following three sets which partition the space $\ncal$:
    \begin{itemize}
    \item solvent and liquid: $\scal(q, \bar q)=\{i \in \ncal \; | \; h_i \leq q^\T [I-A\theta_{\min}] s_i\}$;
    \item solvent but illiquid: $\lcal(q, \bar q)=\{i \in \ncal \; | \; q^\T [I-A\theta_{\min}] s_i  < h_i < \bar q^\T s_i\}$; or
    \item insolvent: $\dcal(q, \bar q)=\{i \in \ncal \; | \; h_i \geq \bar q^\T s_i\}$.
    \end{itemize}
For simplicity of notation, define $\lcal^* := \lcal(q^*,\bar q^*)$.
Let $(q,\bar q) \in \D \mapsto \hat \Gamma(q,\bar q) := \sum_{i \in \lcal(q,\bar q)} \gamma_i(q,\bar q)$ and $(q,\bar q) \in \D \mapsto \hat H(q,\bar q) := \sum_{i \in \lcal(q,\bar q)} h_i$.
We wish to note that $\pdv{\gamma_i(q,\bar q)}{q_k} = \pdv{\gamma_i(q,\bar q)}{\bar q_k} = 0$ for any asset $k$ with prices $(q,\bar q) \in \D$ such that $i \in \scal(q,\bar q) \cup \dcal(q,\bar q)$.

The minimal liquidation condition~\eqref{eq:mlc} implies that:
\[(\bar q - [I - A\theta_{\min}]q)^\T \hat \Gamma(q,\bar q) = \hat H(q,\bar q) - q^\T [I - A\theta_{\min}] \sum_{i \in \lcal(q,\bar q)} s_i\]
for any $(q,\bar q) \in \D$.  Assuming no bank is at the boundary between $\lcal(q,\bar q)$ and either $\scal(q,\bar q)$ or $\dcal(q,\bar q)$ (if so then one-sided derivatives would be required), we are able to determine by implicit differentiation that for any prices $(q,\bar q) \in \D$ and asset $k$
\begin{align}
\label{eq:sensitivity-pf1} [1 - \alpha_k\theta_{\min}]\left(\sum_{i \in \lcal(q,\bar q)} s_{ik} - \hat\Gamma_k(q,\bar q)\right) &= -(\bar q - [I - A\theta_{\min}]q)^\T \pdv{\Gamma^*(q,\bar q)}{q_k}\\
\label{eq:sensitivity-pf2} \hat\Gamma_k(q,\bar q) &= -(\bar q - [I - A\theta_{\min}]q)^\T \pdv{\Gamma^*(q,\bar q)}{\bar q_k}.
\end{align}

By assumption and previous discussions of firm behaviors, $(q,\bar q) \in \D \mapsto \bar F(\Gamma^*(q,\bar q))^\T \hat\Gamma(q,\bar q) + F(\Gamma^*(q,\bar q))^\T [I - A\theta_{\min}] \left(\sum_{i \in \lcal^*} s_i - \hat\Gamma(q,\bar q)\right)$ is strictly decreasing on $(q,\bar q) \in \{(q,\bar q) \in \D \; | \; \lcal(q,\bar q) = \lcal^*\}$.
By differentiation w.r.t.\ the MTMP $q_k$, this implies at the clearing prices $(q^*,\bar q^*)$
\begin{align*}
0 &> \bar F(\Gamma^*(q^*,\bar q^*))^\T \pdv{\Gamma^*(q^*,\bar q^*)}{q_k} + \left(\pdv{\bar F(\Gamma^*(q^*,\bar q^*))}{q_k}\right)^\T \hat\Gamma(q^*,\bar q^*) \\
&\qquad - F(\Gamma^*(q^*,\bar q^*))^\T [I - A\theta_{\min}] \pdv{\Gamma^*(q^*,\bar q^*)}{q_k} + \left(\pdv{F(\Gamma^*(q^*,\bar q^*))}{q_k}\right)^\T[I - A\theta_{\min}]\left(\sum_{i \in \lcal^*} s_i - \hat\Gamma(q^*,\bar q^*)\right)
\end{align*}
for any asset $k$.
That is, for any asset $k$,
\begin{align*}
&\left(\pdv{\bar F(\Gamma^*(q^*,\bar q^*))}{q_k}\right)^\T \hat\Gamma(q^*,\bar q^*) + \left(\pdv{F(\Gamma^*(q^*,\bar q^*))}{q_k}\right)^\T[I - A\theta_{\min}]\left(\sum_{i \in \lcal^*} s_i - \hat\Gamma(q^*,\bar q^*)\right) \\
&\qquad < -\left(\bar F(\Gamma^*(q^*,\bar q^*)) - [I - A\theta_{\min}]F(\Gamma^*(q^*,\bar q^*))\right)^\T \pdv{\Gamma^*(q^*,\bar q^*)}{q_k}\\
&\qquad = -\left(\bar q^* - [I - A\theta_{\min}]q^*\right)^\T \pdv{\Gamma^*(q^*,\bar q^*)}{q_k}.
\end{align*}
Through comparison with~\eqref{eq:sensitivity-pf1}, we are able to conclude that
\begin{equation}\label{eq:sensitivity-pf3}
\begin{split}
&\left(\pdv{\bar F(\Gamma^*(q^*,\bar q^*))}{q_k}\right)^\T \hat\Gamma(q^*,\bar q^*) + \left(\pdv{F(\Gamma^*(q^*,\bar q^*))}{q_k}\right)^\T[I - A\theta_{\min}]\left(\sum_{i \in \lcal^*} s_i - \hat\Gamma(q^*,\bar q^*)\right) \\
&\qquad < [I - A\theta_{\min}]\left(\sum_{i \in \lcal^*} s_i - \hat\Gamma(q^*,\bar q^*)\right).
\end{split}
\end{equation}
Through the same analysis but applying~\eqref{eq:sensitivity-pf2} to the derivatives w.r.t.\ $\bar q_k$, we can also conclude
\begin{equation}\label{eq:sensitivity-pf4}
\begin{split}
&\left(\pdv{\bar F(\Gamma^*(q^*,\bar q^*))}{\bar q_k}\right)^\T \hat\Gamma(q^*,\bar q^*) + \left(\pdv{F(\Gamma^*(q^*,\bar q^*))}{\bar q_k}\right)^\T[I - A\theta_{\min}]\left(\sum_{i \in \lcal^*} s_i - \hat\Gamma(q^*,\bar q^*)\right) \\
&\qquad < \hat\Gamma(q^*,\bar q^*).
\end{split}
\end{equation}

Finally, we consider the vector
\begin{equation*}
v := \left(\begin{array}{c} \left[I - A\theta_{\min}\right]\left(\sum_{i \in \lcal^*} s_i - \hat\Gamma(q^*,\bar q^*)\right) \\ \hat\Gamma(q^*,\bar q^*) \end{array}\right) \geq \vec{0}.
\end{equation*}
By construction of the matrix $W$ and an application of~\eqref{eq:sensitivity-pf3} and~\eqref{eq:sensitivity-pf4}, we find that
\begin{align*}
W^\T v &= \left(\begin{array}{c} \left(\pdv{\bar F(\Gamma^*(q^*,\bar q^*))}{q_1}\right)^\T \hat\Gamma(q^*,\bar q^*) + \left(\pdv{F(\Gamma^*(q^*,\bar q^*))}{q_1}\right)^\T [I - A\theta_{\min}] \left(\sum_{i \in \lcal^*} s_i - \hat\Gamma(q^*,\bar q^*)\right) \\ \vdots \\ \left(\pdv{\bar F(\Gamma^*(q^*,\bar q^*))}{q_m}\right)^\T \hat\Gamma(q^*,\bar q^*) + \left(\pdv{F(\Gamma^*(q^*,\bar q^*))}{q_m}\right)^\T [I - A\theta_{\min}] \left(\sum_{i \in \lcal^*} s_i - \hat\Gamma(q^*,\bar q^*)\right) \\
    \left(\pdv{\bar F(\Gamma^*(q^*,\bar q^*))}{\bar q_1}\right)^\T \hat\Gamma(q^*,\bar q^*) + \left(\pdv{F(\Gamma^*(q^*,\bar q^*))}{\bar q_1}\right)^\T [I - A\theta_{\min}] \left(\sum_{i \in \lcal^*} s_i - \hat\Gamma(q^*,\bar q^*)\right) \\ \vdots \\ \left(\pdv{\bar F(\Gamma^*(q^*,\bar q^*))}{\bar q_m}\right)^\T \hat\Gamma(q^*,\bar q^*) + \left(\pdv{F(\Gamma^*(q^*,\bar q^*))}{\bar q_m}\right)^\T [I - A\theta_{\min}] \left(\sum_{i \in \lcal^*} s_i - \hat\Gamma(q^*,\bar q^*)\right) \end{array}\right) < v.
\end{align*}
Thus an application of, e.g., Theorem 2.1 of \cite{tenraa_2006}, $(I - W)^{-1}$ exists and is given by the Leontief inverse.
\end{proof}

\end{document}